\documentclass[aps,prb,twocolumn,groupedaddress,footinbib]{revtex4-2}

\usepackage{amsmath}
\usepackage[dvips]{color}
\usepackage[utf8]{inputenc}
\usepackage[american]{babel}
\usepackage{color}
\usepackage{graphicx}
\usepackage{epstopdf}
\usepackage{epsfig}
\usepackage{float}
\usepackage[colorlinks,linkcolor=blue,citecolor=blue,urlcolor=blue]{hyperref}
\usepackage{comment}
\usepackage{amssymb}
\usepackage{multirow}
\usepackage[table]{xcolor}
\usepackage[normalem]{ulem}

\allowdisplaybreaks

\def\k{{ {\mathbf k} }}

\def\q{{ {\mathbf q} }}

\usepackage{float}

\begin{document}

	\title{Displaced Drude peak from $\pi$-ton vertex corrections} 
	
	\author{J. Krsnik$^{1,2}$} 
	\email{juraj.krsnik@tuwien.ac.at} 
	\author{O. Simard$^{3,4,5}$, P. Werner$^4$, A. Kauch$^1$, and K. Held$^1$}

	\affiliation{$^{1}$Institute of Solid State Physics, TU Wien, 1040 Vienna, Austria,\\
 $^{2}$Department for Research of Materials under Extreme Conditions, Institute of Physics, HR-10000 Zagreb, Croatia\\
		$^{3}$CPHT, CNRS, Ecole Polytechnique, Institut Polytechnique de Paris, Route de Saclay, 91128 Palaiseau, France\\
		$^{4}$Department of Physics, University of Fribourg, 1700 Fribourg, Switzerland\\
        $^{5}$Collège de France, 11 place Marcelin Berthelot, 75005 Paris, France}
	
	\begin{abstract}
		Correlated electron systems  often show strong  bosonic fluctuations, e.g., of antiferromagnetic nature, around  a large wave vector such as $\q=(\pi,\pi\ldots)$. These fluctuations can give rise to vertex corrections to the  optical conductivity through the  (transversal) particle-hole channel, coined $\pi$-ton contributions. Previous numerical results differed qualitatively on how such vertex corrections alter the optical conductivity. Here, we clarify that  $\pi$-ton vertex corrections lead to a displaced Drude peak for correlated metals.
 The proximity and enhancement of the effect when approaching  a phase transition of, e.g., antiferromagnetic nature 
   can be utilized  for  discriminating $\pi$-tons in experiments from other physics leading to a displaced Drude peak. 
	\end{abstract}
	
	\maketitle 
	
	\section{Introduction}
	
	Optical probes are important tools in condensed matter physics for studying the electronic properties of materials.  The response of an electrical current to an external dynamic electric field is described by the optical conductivity. For a generic metallic system, it is peaked at zero frequency as predicted in the classical Drude theory \cite{drude1,drude2}, while for semiconducting or insulating systems an optical gap in its spectrum appears, since a nonzero photon energy is needed to excite electrons across the band gap. 
	In the simplest theory, both of these spectra can be described by the uncorrelated propagation of an electron and a hole, corresponding in the diagrammatic language to the bubble term. In certain cases, however, correlations between the excited electron and hole lead to novel physical phenomena, requiring a proper treatment of vertex correction contributions. The prime examples are excitons in semiconductors
 \cite{frenkel1931,wannier1937} 
 and weak localization in disordered systems \cite{abrahams1979,gorkov1979,gotze1979,altshuler1980}. In the former case, the electron and hole form a bound state giving rise to excitonic peaks in the optical gap, while the suppression of the DC conductivity due to the destructive interference of the electron wave function occurs in the latter case.
	
	Naturally, the question arises: do vertex corrections play  an equally  important role in shaping an optical conductivity spectrum of correlated electron systems? This long-standing fundamental question has captured the interest of the community for several decades \cite{clarke1993,maebashi1997,essler2001,wrobel2002,jeckelmann2003,kontani2006,lin2009,bergeron2011,chubukov2014,kokalj2017,maslov2017,vucicevic2019,huang2019}. However, it was only until quite recently that a convenient classification and an identification of the important class of vertex corrections in correlated electron systems was carried out \cite{pudleiner2019,kauch2020,astleithner2020}. This was possible due to the recent methodological advances in using the parquet equations \cite{bickers2004,gang2016,gangli2016} within the dynamical vertex approximation \cite{kusunose2006,toschi2007,katanin2009} and the parquet approximation \cite{bickers2004}. This allows for classifying and studying the vertex corrections according to the two-particle reducibility. As a result, it was found that the dominant vertex corrections in prototypical models of strongly correlated electrons  are those in the transversal particle-hole $(\overline{ph})$ channel. This is because the   $\overline{ph}$ channel can pick up bosonic fluctuations at an arbitrary wave vector, even though the transfer momentum of the photon is zero. Specifically, strong antiferromagnetic (AFM) or charge density wave fluctuations at $\mathbf{k}-\mathbf{k}' \approx (\pi,\pi,...)$  enter the optical conductivity via the  $\overline{ph}$ channel, and have been coined $\pi$-ton vertex corrections \cite{kauch2020}.
 Excitons, on the other hand, emerge from the particle-hole ($ph$) channel, and weak localization corrections from the particle-particle ($pp$) channel.
	
	In Refs.~\cite{pudleiner2019,kauch2020,astleithner2020}, $\pi$-ton vertex corrections were studied for several correlated models both in the insulating and in the metallic phases. In the insulating cases, they were reported to shift the optical gap, while in the metallic phases, a renormalization of the Drude peak was observed, but also the displaced Drude peak profile of the total optical conductivity in the case of the metallic phase of the Falicov-Kimball model \cite{falicov1969}. Soon after these numerical studies, simplified random phase approximation (RPA) calculations of the $\pi$-ton vertex corrections were performed \cite{simard_eq,simard_noneq,worm2021}, with the aim of studying in more detail their characteristics in the weakly correlated regime of the Hubbard model. While in Ref.~\cite{worm2021} only a temperature dependent sharpening and broadening of the Drude peak was reported, in Refs.~\cite{simard_eq,simard_noneq} it was argued that an additional $\pi$-ton peak may arise next to an existing Drude peak. Such striking differences could have originated from several sources, such as the dimensionality of a system, i.e., one-dimensional (1D) in Refs.~\cite{simard_eq,simard_noneq} and two-dimensional (2D) in Ref.~\cite{worm2021}, Hubbard bands present in Refs.~\cite{simard_eq,simard_noneq}, but not in Ref.~\cite{worm2021}, or perhaps the pitfalls of the analytic continuation of the optical spectra to real frequencies \cite{simard_eq}.

	In this paper, we try to reconcile the previously conflicting results by evaluating the $\pi$-ton vertex corrections within the RPA using two different approaches. Our main finding is that irrespective of dimensionality, the presence of Hubbard bands, and the use of an analytic continuation, both approaches agree on the qualitative frequency dependence of the $\pi$-ton vertex corrections, which eventually yield displaced Drude peaks in the total optical conductivity.

	A displaced Drude peak is not an uncommon phenomenon and has been experimentally observed for decades \cite{rozenberg1995,puchkov1995,tsvetkov1997,wang1998,osafune1999,takenaka1999,lupi2000,kostic1998,lee2002,takenaka2002,santandersyro2002,takenaka2003,wang2003,hussey2004,wang2004,takenaka2005,jonsson2007,kaiser2010,jaramillo2014,biswas2020,pustogow2021,uzkur2021,uykur2022}. As already mentioned, it naturally arises as a consequence of the weak localization effects in disordered systems. However, its unambiguous presence in correlated materials calls for a better or alternative understanding of the phenomenon, with several theories already put forth \cite{luca2017,fratini2021,rammal2023}. While in Ref.~\cite{luca2017} it is explained in terms of the hydrodynamics of short-range quantum critical fluctuations of incommensurate density wave order, in   Refs.~\cite{fratini2021,rammal2023} it is argued that quantum localization corrections may arise due to the slow phononic fluctuations. Here, we show that the strong AFM fluctuations through the $\pi$-ton vertex corrections may as well lead to a displaced Drude peak. This represents a microscopic theory of the displaced Drude peak formation in clean correlated electron systems, where the bosonic fluctuation emerges from intrinsic many-body electron interactions.
	
	The outline of the paper is as follows:
	In Section \ref{sec:model}, we briefly introduce the Hubbard model and how optical conductivities are calculated. Further  details are given for  the three different parameter sets studied in Section~\ref{ssec:selfenergy}, 
	and the RPA calculation of vertex corrections in Section~\ref{ssec:vertex}. 
	Section~\ref{sec:results} presents our results. Here, in Section \ref{ssec:dDrude} we demonstrate the occurrence of a displaced Drude peak; in Section~\ref{ssec:fate_vertex} we employ an adaptive  $\nu$- and $\k$-space integration and show results in the immediate vicinity of the antiferromagnetic phase transition; and in Section~\ref{ssec:Tdep}
	we show the temperature dependence of the peak position and height of the displaced Drude peak.
	Finally, in Section~\ref{ssec:pastresults}, we discuss why previous calculations did not see the displaced Drude peak, before summarizing our results in Section~\ref{sec:conclusion}.

	\section{Model and methods}
	\label{sec:model}
	
	We consider a single-orbital Hubbard model on a D-dimensional hypercubic lattice
	
	\begin{equation} \label{eq:hubbard_hamiltonian}
		\hat{\mathcal{H}} = -t \sum_{\left\langle ij\right\rangle\sigma }^{} \hat{c}_{i\sigma}^\dagger \hat{c}_{j\sigma} + U \sum_{i}^{}\hat{n}_{i\uparrow}\hat{n}_{i\downarrow}\:,
	\end{equation}
	where $\hat{c}_{i\sigma}^\dagger (\hat{c}_{i\sigma})$ denotes the fermionic creation (annihilation) operator for an electron at lattice site $i$ with spin $\sigma$, and $\hat{n}_{i\sigma} = \hat{c}_{i\sigma}^\dagger\hat{c}_{i\sigma}$ is a number operator. We take into account hoppings only between nearest neighbors, indicated by $\left\langle ij \right\rangle$ in the sum over lattice sites (each pair of neighboring sites is counted once in the sum). In all our calculations, we set hopping $t\equiv 1 $ as the unit of energy, as well as $\hbar \equiv 1$, $k_B \equiv 1$, electric charge $e\equiv 1$, and lattice constant $a\equiv 1$. We keep the value of the local interaction $U\leq 2$ and hence focus on the weakly correlated metallic state. Finally, we consider the half-filled case for which we anticipate the presence of strong antiferromagnetic fluctuations.

\begin{figure}
		\centering
		\includegraphics[width=.48\textwidth]{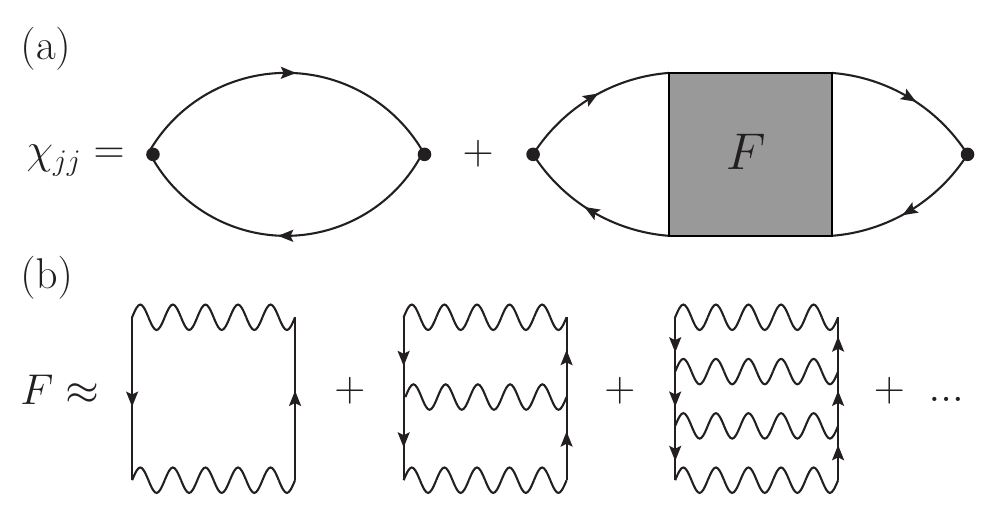}
		\renewcommand{\figurename}{Fig.}
		\caption{ (a) Diagrammatic representation of the current-current correlation function $\chi_{jj}$ with bubble (left) and vertex contribution (right).  (b) Diagrammatic representation of the $\pi$-ton vertex corrections in the $\overline{ph}$ channel within the RPA. Here, the solid lines represent the fermion Green's functions $G$, the wavy lines the Hubbard interaction $U$, $F$ is the vertex function, while light-fermion vertices are denoted by solid circles.}
		\label{fig:diagram}
	\end{figure}

	We are interested in calculating the  optical conductivity $\sigma(\omega)$ for the model Hamiltonian Eq.~\eqref{eq:hubbard_hamiltonian}. We obtain it from the current-current correlation function $\chi_{jj}$, which  we calculate in turn  using two distinct approaches. Within the first approach, here called the real-axis approach, the current-current correlation function $\chi_{jj}(\omega)$ is calculated directly on the real-frequency axis using the expressions derived in Ref.~\cite{worm2021}. The optical conductivity is then simply obtained as $\sigma(\omega) = \text{Im}\chi_{jj}(\omega)/\omega$, while its static limit, that is the DC conductivity, $\sigma_{DC} = \lim\limits_{\omega\to 0}\sigma(\omega)$, is handled by extrapolating finite frequency values to the zero frequency limit.
	In the second approach, here referred to as the imaginary-axis approach, the current-current correlation function $\chi_{jj}(i\omega_n)$ is instead calculated on the imaginary-frequency axis as described in Ref.~\cite{simard_eq}. To get the real frequency optical conductivity spectrum, the use of analytic continuation is thus needed, which we perform using the ana\_cont package \cite{ana_cont}. Common to both approaches is that $\chi$ and $\sigma$ are separated into the bubble (BUB) and the vertex (VERT) contribution, Fig.~\ref{fig:diagram}(a). For the bubble contribution, only the knowledge of the fermion Green's functions $G$ is needed, while for the vertex contribution, the vertex function $F$ is additionally required. In the following, we outline the ideas behind modeling both $G$ and $F$, such that $\pi$-ton vertex corrections are treated within the RPA.
	
	\subsection{Models for the self-energy}
	\label{ssec:selfenergy}
	
	The aforementioned real-axis approach allows for any fermion Green's function $G$,
	as long as the self-energy $\Sigma$, with or without momentum dependence, is known on the real-frequency axis.
	Since neglecting the momentum and retaining only the frequency dependence of the self-energy in the spirit of dynamical mean-field theory (DMFT) \cite{georges1996} is sufficient to capture the effects of Hubbard bands on the $\pi$-ton vertex corrections, we do not consider the effects of a possible momentum dependence of $\Sigma$  in the following. In particular, in order to address the influence of Hubbard bands as well as dimensionality effects on the $\pi$-ton vertex corrections, we define three sets of parameters $S_{\text{DMFT}}$, $S_{\text{1D}}$, and $S_{\text{2D}}$ to model the fermion Green's functions as follows.
	
	\subsubsection{Parameter set $S_{\text{DMFT}}$}
	
	In the spirit of Ref.~\cite{simard_noneq}, the parameter set $S_{\text{DMFT}}$ regards a 1D lattice with $U = 2$, and the DMFT self-energy obtained with iterated perturbation theory (IPT) \cite{kajueter1996}.  Such a choice of self-energy is convenient for the calculation of the current-current correlation function within the imaginary-axis approach \cite{simard_eq}. Out of the several temperatures considered in Ref.~\cite{simard_noneq}, we mainly focus on the case $\beta = 12.5$. For the corresponding temperature $T=\beta^{-1}$, it has been reported that the $\pi$-ton vertex corrections show a distinct broad peak at $\omega\approx0.35$ \cite{simard_noneq}. We would like to note that the DMFT phase diagram in 1D resembles that of higher dimensional systems and does not capture the presence of a Mott gap for any $U>0$, however small,  at half-filling. 
	Only for $U$ of the order of the bandwidth does a Mott gap open. Hence, our 1D results should primarily serve as a proxy for the behavior of the $\pi$-ton vertex corrections in higher dimensional systems or slightly away from the half-filled case.
	
	\subsubsection{Parameter set $S_{\text{1D}}$}
	
	Motivated by studying the effects of Hubbard bands on the $\pi$-ton vertex corrections, we introduce the second parameter set $S_{\text{1D}}$, referring again to the 1D lattice with $U=2$, but now with the simplified frequency independent self-energy
	
	\begin{equation} \label{eq:const_self_energy}
		\Sigma = - i \Delta(T)\:,\quad  \Delta(T)=0.1547+1.637\;T^2\;.
	\end{equation}
	This form of self-energy can be rationalized by the ever-present impurity scattering and resembles the Fermi-liquid-like temperature behavior \cite{coleman2015}. The choice of parameter values entering $\Delta(T)$ is motivated in Ref.~\cite{worm2021}.
	
	The difference between $S_{\text{DMFT}}$ and $S_{\text{1D}}$ is that the former involves both the quasiparticle and Hubbard bands, while the parameter set $S_{\text{1D}}$ involves only the former. By comparing the optical conductivity between both sets, we can thus single out features that result from the presence of Hubbard bands.
	
	\subsubsection{Parameter set $S_{\text{2D}}$}
	
	Our third set of parameters $S_{\text{2D}}$ relates to the 2D square lattice with $U = 1.9$ and the same self-energy as in Eq.~\eqref{eq:const_self_energy}. The point of introducing it is to study differences between the $\pi$-ton vertex corrections in 1D and 2D cases. For both parameter sets $S_{\text{1D}}$ and $S_{\text{2D}}$, we calculate the current-current correlation function using the real-axis approach.
	
	\subsection{Vertex function}
	\label{ssec:vertex}
	
	The evaluation of the vertex contribution to the current-current correlation function requires the knowledge of the full density component of the two-particle vertex $F^{kk'q}_d$ \cite{gang2016,rohringer2018}. Following arguments in Ref.~\cite{kauch2020}, we, however, only focus on the vertex contributions $F^{kk'q}_{d,\overline{ph}}$ coming from the $\overline{ph}$ channel - $\pi$-ton vertex corrections, which are supposed to be dominant in the Hubbard model. In order to be able to evaluate the corresponding $\pi$-ton vertex contributions to the current-current correlation function within the real-axis approach of Ref.~\cite{worm2021} and/or imaginary-axis approach of Ref.~\cite{simard_eq}, we further assume that the vertex function depends only on one transfer momentum and energy, $F^{kk'q}_{d,\overline{ph}}\equiv F^{k-k'}_{d,\overline{ph}}$.  Here $k=({\mathbf k},\nu)$ denotes a combined momentum and frequency index. Generally, such contributions may still be quite complicated, but diagrammatically they can be represented as vertical ladders in terms of the irreducible vertex $\Gamma_{\overline{ph}}$. Following further the modeling approach of the vertex function described in Refs.~\cite{worm2021,simard_eq}, we focus on the $\uparrow\uparrow\downarrow\downarrow\;\equiv\;\uparrow\downarrow$ spin component and take for the irreducible vertex $\Gamma_{\overline{ph}} = -U$. Namely, the building blocks of our vertical ladders are the interaction $U$ and the Lindhard function $\chi^0_{q}=-\frac{1}{\beta}\sum_{k}^{}G_{k}G_{k+q}$, see Fig.~\ref{fig:diagram}(b). This series of vertical ladders can be summed up to infinite order to finally yield the RPA version of the $\pi$-ton vertex function
	
	\begin{equation} \label{eq:RPA_vertex}
		F^{RPA}_{\overline{ph}, k-k'} = \frac{U^2\chi^0_{k-k'}}{1-U\chi^0_{k-k'}} = U^2\chi^{RPA}_{k-k'}\;,
	\end{equation}
	where $\chi^{RPA}_{k-k'}$ is the RPA magnetic susceptibility, which we use in all our calculations of the vertex contribution to the current-current correlation function.
	Note that $F^{RPA}_{\overline{ph}, k-k'}$ is fully determined by $G$ (respectively $\Sigma$), $U$, and $T$.

	\subsubsection{Paramagnetic-antiferromagnetic transition}
	
	For a given parameter set $S_i$, there exists a critical temperature $T_c = \beta^{-1}_c$ for which the vertex in Eq.~\eqref{eq:RPA_vertex} diverges. Within our model, this divergence signals the  paramagnetic-to-antiferromagnetic phase transition. Due to the RPA treatment of the vertex, the transition appears at nonzero temperature even for 1D and 2D systems, violating the Mermin-Wagner theorem \cite{mermin1966}. Hence, our results should again serve only as a proxy for the behavior of the $\pi$-ton vertex corrections near the transition boundary. Once the critical temperature is determined, we place ourselves within the paramagnetic metallic state with $T \geq T_c$. Then we lower the temperature towards the critical one, thereby enhancing the antiferromagnetic fluctuations and the overall effect of the $\pi$-ton vertex corrections.
	
	There is no guarantee that the $T_c$, determined as the temperature at which the RPA vertex in Eq.~\eqref{eq:RPA_vertex} diverges, matches the Néel temperature $T_N$ obtained by the IPT solver. Indeed, while the two values are quite close for $U\leq 1$ \cite{simard_eq}, for larger values of $U$ the discrepancies between the two become larger. Therefore, a renormalized $U_r$ was introduced in Ref.~\cite{simard_eq}
	
	\begin{equation} \label{eq:RPA_vertex_renormalized}
		F^{RPA,r}_{\overline{ph}, k-k'} = \frac{U^2\chi^0_{k-k'}}{1-U_r\chi^0_{k-k'}}\;,
	\end{equation}
	to push $T_c$ towards $T_N$. For the parameter set $S_{\text{DMFT}}$ this is achieved by taking $U_r = 1.33$, yielding $\beta_c\approx \beta_N \approx 20$ \cite{simard_noneq}.
	
	Regarding parameter sets $S_{\text{1D}}$ and $S_{\text{2D}}$, we keep $U_r = U$, resulting in $\beta_c\approx 23$ and $\beta_c \approx 19$ for the former and the latter parameter set, respectively. It now becomes apparent that all parameter sets are chosen such that their $\beta_c$ are roughly similar.
	
	\subsubsection{Ornstein-Zernike form of the vertex function}
	
	Close to the transition boundary, in the presence of strong antiferromagnetic fluctuations, the magnetic susceptibility and thus our $\pi$-ton vertex function can be well approximated by the Ornstein-Zernike correlation function of the form \cite{hertz1976,millis1990,lohnezsen2007}
	
	\begin{equation} \label{eq:oz_form}
		F^{\text{OZ}}_{\mathbf{q},\omega} = \frac{A}{\xi^{-2}+\left( \mathbf{q}-\mathbf{Q}\right)^2 - i\lambda\omega }\;.
	\end{equation}
	Here, $\xi$ is the correlation length of the antiferromagnetic fluctuations, while $A$ and $\lambda$ represent their effective coupling strength to fermions and the damping rate, respectively. For our half-filled Hubbard model $\mathbf{Q}=\left(\pi,\pi,... \right) $.  One of the advantages of working with the Orsntein-Zernike vertex function is that it can be readily calculated on the fly for any $\mathbf{q}$ and $\omega$, which we exploit to adaptively integrate the vertex contribution to the current-current correlation function within the real-axis approach close to the transition boundary. For practical purposes, in all our actual calculations involving Eq.~\eqref{eq:oz_form} for the vertex function, we use an empirically more robust formula with $\left( \mathbf{q}-\mathbf{Q}\right)^2 \to 4\sum_{i}^{}\sin^2\left( \frac{q_i-Q_i}{2}\right) $ \cite{schafer2021}.
	
		\begin{figure*}
		\centering
		\includegraphics[width=.99\textwidth]{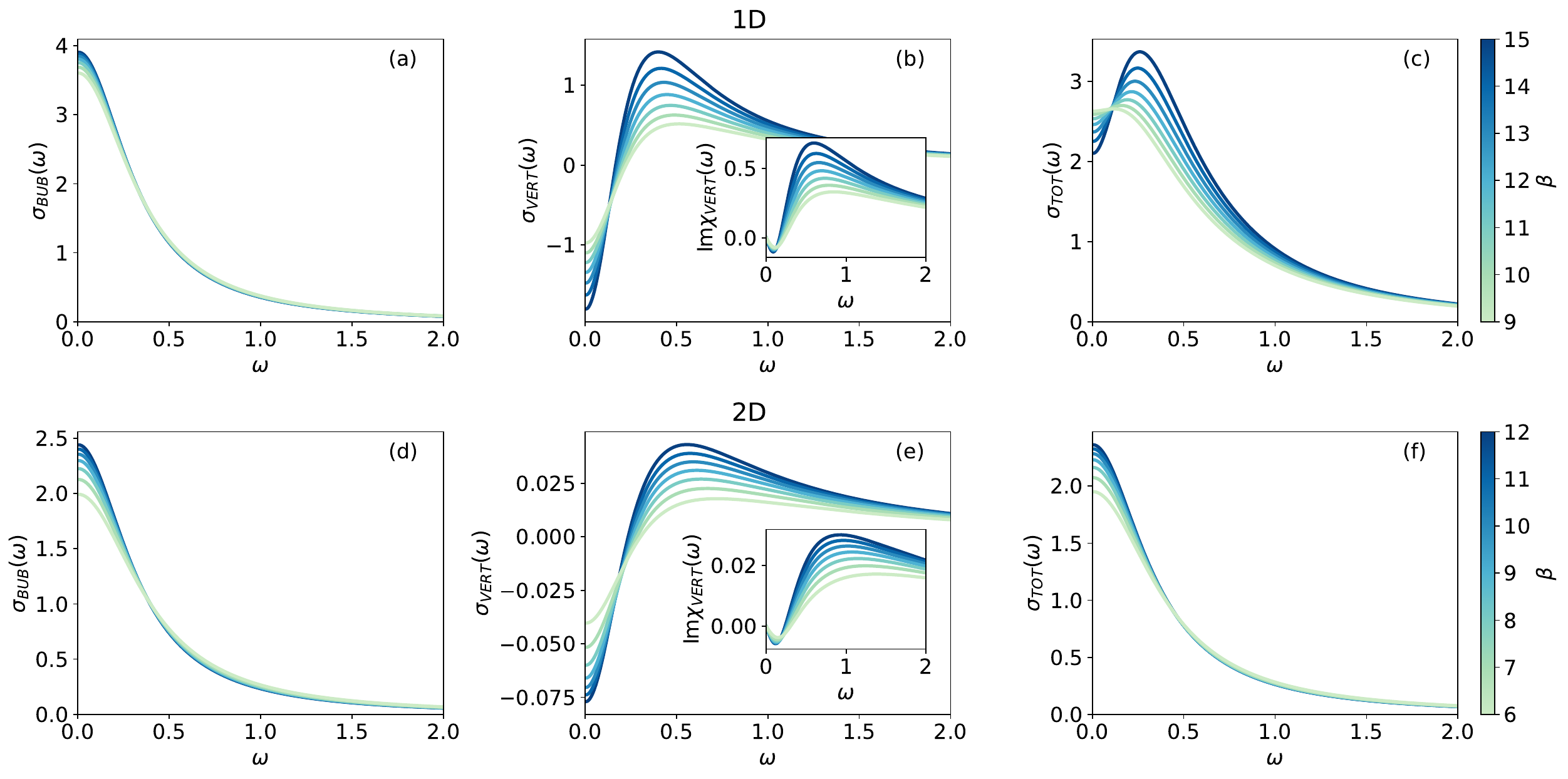}
		\renewcommand{\figurename}{Fig.}
		\caption{Bubble (a,d), RPA $\pi$-ton vertex (b,e), and total contribution  (c,f) to the optical conductivity for the parameter sets $S_{\text{1D}}$ (a-c) and $S_{\text{2D}}$ (d-f) and several temperatures $T=1/\beta$. The insets show the corresponding RPA $\pi$-ton vertex contributions to the current-current correlation function.}
		\label{fig:optical_conductivity_converged_RPA}
	\end{figure*}
	
	\section{Results}
	\label{sec:results}
	
	\subsection{Displaced Drude peak from $\pi$-ton vertex corrections} \label{subsec:real_axis}
	\label{ssec:dDrude}
	
	Optical conductivities calculated directly on the real-frequency axis using Eqs.~(5), (7), (8a), and (8b) of Ref.~\cite{worm2021} and the RPA vertex function in Eq.~\eqref{eq:RPA_vertex} for the parameter sets $S_{\text{1D}}$ and $S_{\text{2D}}$ are shown in Figs.~\ref{fig:optical_conductivity_converged_RPA}(a-c) and (d-f), respectively. 
	The corresponding summations over the Brillouin zone are performed on the grid sizes $N_k = 200$ and $N_k=40 \times 40$ for the 1D and 2D cases, respectively.
	The remaining integrals over frequencies are evaluated using the trapezoidal rule on the grids with $N_\nu = 3193$ points in the range $[-8,8]$ in 1D and on the  grids with $N_\nu = 1593$ points in the range $[-7.6,7.6]$ in the 2D cases. These choices of summation/integration grids were thoroughly checked to produce converged results for all the temperatures considered in Fig.~\ref{fig:optical_conductivity_converged_RPA}. For lower temperatures approaching the critical temperature $T_c$, the convergence of the $\pi$-ton vertex contribution becomes numerically more and more demanding. We discuss this issue in more detail in the next  Subsection \ref{ssec:fate_vertex}.

	As expected, the bubble contribution to the optical conductivity $\sigma_{BUB}$, shown in Figs.~\ref{fig:optical_conductivity_converged_RPA}(a,d), gives the Drude peak in both 1D and 2D cases. The corresponding widths and maxima of these peaks are governed by the fermion scattering rate, which is, with our choice of parameter sets $S_{\text{1D}}$ and $S_{\text{2D}}$, given simply by $\tau\left( T\right)=\frac{1}{2\Delta(T)}$. Correspondingly, the magnitudes of $\sigma_{BUB}$ in both Figs.~\ref{fig:optical_conductivity_converged_RPA}(a,d) are roughly similar.
	
	On the other hand, while the $\pi$-ton vertex contributions to the optical conductivity, $\sigma_{VERT}$ in Figs.~\ref{fig:optical_conductivity_converged_RPA}(b,e), show a  qualitatively similar behavior in both cases (1D and 2D), their magnitude is up to two orders of magnitude larger in the former case, at least for the considered temperature ranges. Taking that into account together with the peculiar frequency dependence of the $\pi$-ton vertex contributions, the resulting total optical conductivity, $\sigma_{TOT}=\sigma_{BUB}+\sigma_{VERT}$, in Figs.~\ref{fig:optical_conductivity_converged_RPA}(c,f) apparently exhibits quite different structures depending on the dimension. In particular, the results in Fig.~\ref{fig:optical_conductivity_converged_RPA} suggest that the $\pi$-ton vertex corrections suppress the optical conductivity at low frequencies  and develop a broad maximum at some intermediate frequency, while for larger frequencies they asymptotically decay to zero. If the magnitude of such vertex corrections is large, as is the case in our 1D calculations, then the sum of the bubble and the $\pi$-ton vertex contributions results in a displaced Drude peak, as can be seen in Fig.~\ref{fig:optical_conductivity_converged_RPA}(c). In contrast, if the magnitude of such vertex corrections is small compared to the bubble contribution, then the sum results in a broadened Drude peak, as in our  2D calculations, see Fig.~\ref{fig:optical_conductivity_converged_RPA}(f). In the following, we get even closer to the transition boundary, enhancing thus AFM fluctuations and $\pi$-ton vertex contributions, and present further arguments in favor of the formation of the displaced Drude peak in both cases, 1D and 2D.
	
	\subsection{$\pi$-ton vertex corrections close to the transition boundary}  \label{ssec:fate_vertex}
	
	Close to the paramagnetic-antiferromagnetic transition boundary the $\pi$-ton vertex function resembles the Ornstein-Zernike form of Eq.~\eqref{eq:oz_form}. As the temperature is lowered towards the critical temperature, the correlation length $\xi$ of the antiferromagnetic fluctuations increases, resulting in the confinement of the $\pi$-ton vertex function around momentum $\mathbf{Q}=\left(\pi,\pi,... \right) $ and frequency $\omega=0$. To be specific, the widths of the Ornstein-Zernike function around these points read
	
	\begin{equation}
		\Gamma_q \sim \frac{1}{\xi}\;,\quad \Gamma_\omega \sim \frac{1}{\lambda\xi^2}\;,
	\end{equation}
	which are getting narrower as the transition boundary is approached.
	
	From the computational point of view, to resolve such fine momentum and frequency features of the $\pi$-ton vertex function we would need to use sufficiently dense grids, $\Gamma_q\gg N_k^{-1}$ and $\Gamma_\omega\gg N_\nu^{-1}$. This quickly becomes a numerical bottleneck if equidistant grids are used, especially in the 2D case, which involves four momentum and two frequency summations/integrations in Eqs.~(7), (8a), and (8b) of Ref.~\cite{worm2021}. For that reason, we focus on the parameter set $S_{\text{1D}}$, and adjust the multi-dimensional adaptive integration package cubature \cite{cubature} to our needs, in order to   evaluate the $\pi$-ton vertex contributions to the optical conductivity all the way down to the transition boundary.
	
	Adaptive integration requires computing the integrand on the fly, which is inconvenient with the RPA $\pi$-ton vertex function since it involves momentum and frequency summations at each call. For this reason and since the Ornstein-Zernike form close to the critical temperature  approximates the $\pi$-ton vertex function well, we first calculate the RPA $\pi$-ton vertex function in Eq.~\eqref{eq:RPA_vertex} using dense momentum and frequency grids, and then fit it with the Ornstein-Zernike form in Eq.~\eqref{eq:oz_form} to obtain $A$, $\xi$, and $\lambda$ for a given temperature. In this way, we have the vertex function $F^{\text{OZ}}_{\mathbf{q},\omega}$ as a simple function that can be conveniently used for adaptive integration of the $\pi$-ton vertex corrections according to Ref.~\cite{worm2021}.

	\renewcommand{\arraystretch}{1.25}
	\begin{table}[h!]
		\centering
		\begin{tabular}{|>{\centering\arraybackslash}  m{1.35cm}||>{\centering\arraybackslash}m{1.35cm}| >{\centering\arraybackslash}m{1.35cm} | >{\centering\arraybackslash}m{1.35cm} |} 
			\hline
			& $A $ & $\lambda $& $\xi $\\ 
			\hline
			$\beta = 22$ & 0.501 &	 0.214 &	 37.784 \\ 
			\hline
			$\beta = 21$ & 0.514 &	 0.219 &	 25.976\\
			\hline
			$\beta = 20$ & 0.529 &	 0.228 &	 20.383 \\
			\hline
			$\beta = 19$ & 0.545&	 0.229 &	 16.880\\ 
			\hline
			$\beta = 18$ & 0.564 &	 0.235 &	 14.376 \\ 
			\hline
			$\beta = 17$ &0.586&	 0.242 &	 12.447 \\ 
			\hline
			$\beta = 16$ &0.610 &	 0.249 &	 10.882\\ 
			\hline
		\end{tabular}
		\caption{Ornstein-Zernike parameters $A$, $\lambda$, and $\xi$ extracted by fitting the Ornstein-Zernike form in Eq.~\eqref{eq:oz_form} to the RPA $\pi$-ton vertex function in Eq.~\eqref{eq:RPA_vertex} for different temperatures.}
		\label{table:oz-parameters}
	\end{table}
	\renewcommand{\arraystretch}{1}
	
	The values of the Ornstein-Zernike parameters for the one-dimensional parameter set $S_{\text{1D}}$ and different temperatures obtained by fitting Eq.~\eqref{eq:oz_form} to the RPA $\pi$-ton vertex function are given in Table~\ref{table:oz-parameters}, while the corresponding $\pi$-ton vertex contributions to the optical conductivity obtained with the adaptive integration are shown in Fig.~\ref{fig:OZ_results}. More details on the fitting procedure are given in Appendix~\ref{app:oz_fit}.
	Evidently, the $\pi$-ton vertex contributions continue to show a similar qualitative trend as in the RPA case all the way down to the critical temperature. As expected, their magnitude grows larger as the transition boundary is approached due to the enhanced antiferromagnetic fluctuations, leading to an increasing suppression of the DC conductivity and an increasingly pronounced maximum at finite frequencies.
	
	\begin{figure}
		\includegraphics[width=.48\textwidth]{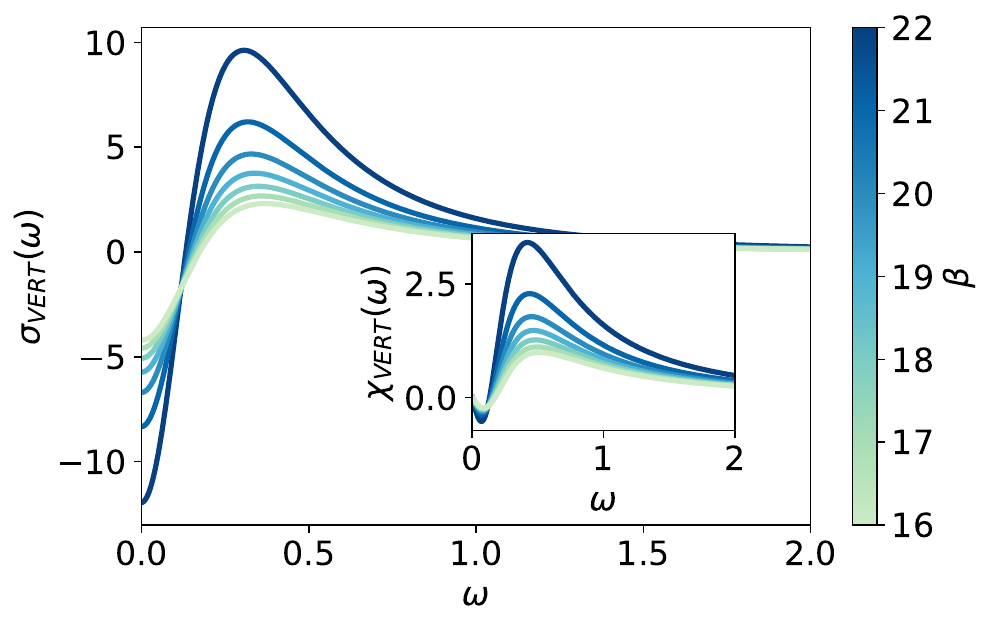}
		\renewcommand{\figurename}{Fig.}
		\caption{$\pi$-ton vertex corrections to the optical conductivity and the current-current correlation function (inset) calculated with the Ornstein-Zernike vertex function and an adaptive integration for temperatures approaching  the critical temperature, $T_c^{-1}=\beta_c \approx 23$, for the parameter set $S_{\text{1D}}$.}
		\label{fig:OZ_results}
	\end{figure}
	
	Two important conclusions can be drawn from the results of the adaptive integration in Fig.~\ref{fig:OZ_results}. The first is that we can anticipate the displaced Drude peak in the optical conductivity closer to the transition boundary where the AFM fluctuations are greatly enhanced also in the 2D case. Achieving convergence of 2D calculations at low temperatures is quite challenging even with adaptive integration, but based on all our converged results it is reasonable to assume that the overall shape of the $\pi$-ton vertex contributions would continue to follow a similar trend as in the 1D case. The second implication is that at a certain point our approximations will fail and one would need to revisit them both from the point of view of the fermion Green's function, $G$, and the vertex function, $F$. It can be already inferred from Fig.~\ref{fig:OZ_results} that the magnitude of the $\pi$-ton vertex contributions at low frequencies may exceed that of the bubble, leading to unphysical negative spectral weight in the total optical conductivity. The intention of the present paper is, however, to point out that the $\pi$-ton vertex corrections generally lead to a displaced Drude peak, not to properly study all of their quantitative features. The latter requires more sophisticated calculations beyond RPA which are, to the best of our knowledge, not feasible with the accuracy - dense $\nu$- and $\k$-meshes - and real frequency calculations required here.
	
	\subsection{Temperature dependence of the displaced Drude peak frequency and height}
	\label{ssec:Tdep} 
	
\begin{figure}
		\centering
		\includegraphics[width=.48\textwidth]{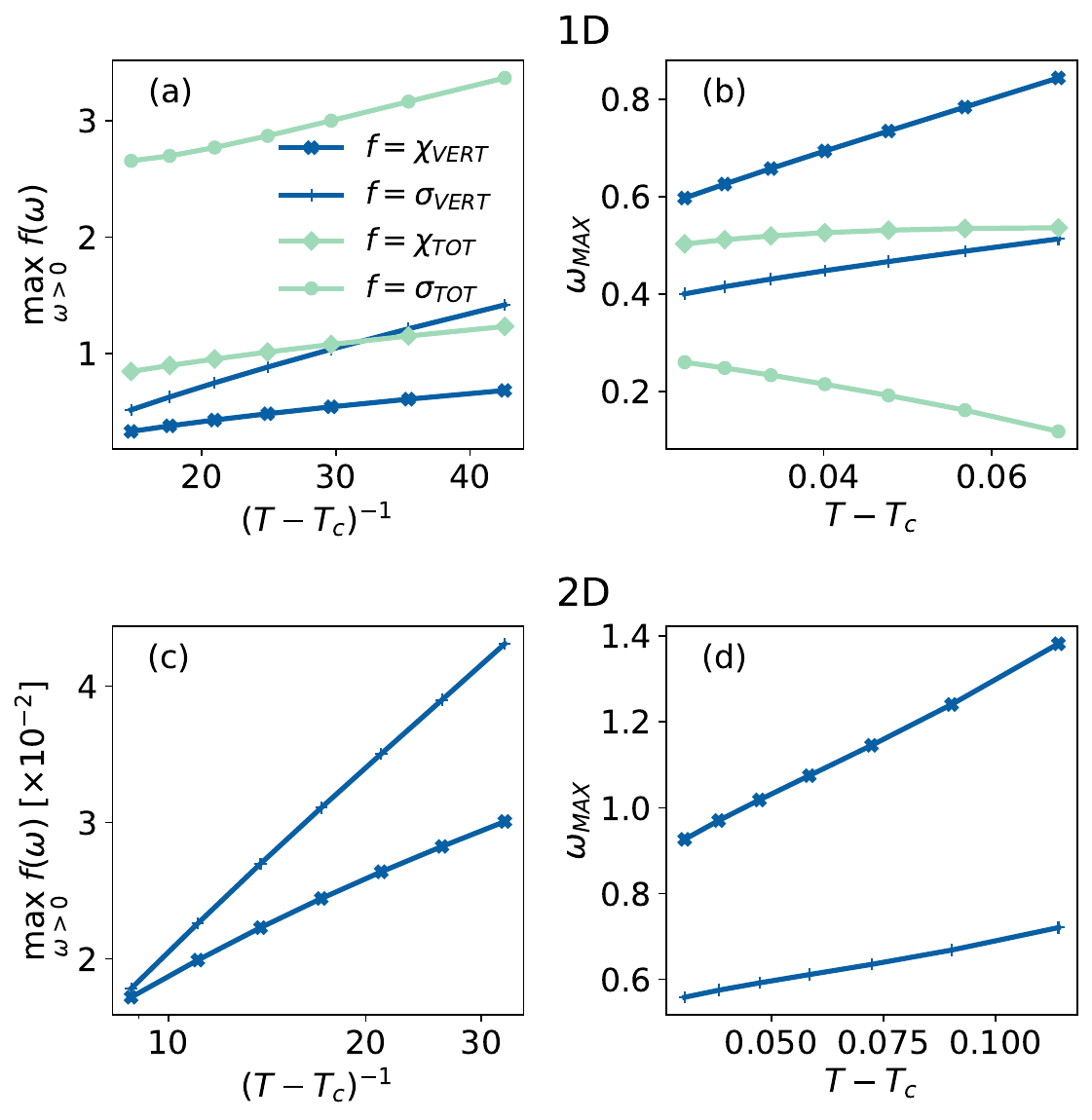}
		\renewcommand{\figurename}{Fig.}
		\caption{(a,c) Temperature dependence of the maxima of the current-current correlation function ($f=\chi$) and the optical conductivity ($f=\sigma$) shown separately for the vertex (VERT; blue lines) and total contribution (TOT; green lines). (b,d) Frequencies $\omega_{MAX}$ associated with the maxima. The top panels (a,b) show the 1D $S_{\text{1D}}$ and the bottom panels (c,d) the 2D $S_{\text{2D}}$ results.}
		\label{fig:max_temp_dependence}
	\end{figure}

	The displaced Drude peak has been experimentally observed in a great variety of compounds, ranging from cuprates, and transition metal oxides to organic conductors and Kagome metals \cite{rozenberg1995,puchkov1995,tsvetkov1997,wang1998,osafune1999,takenaka1999,lupi2000,kostic1998,lee2002,takenaka2002,santandersyro2002,takenaka2003,wang2003,hussey2004,wang2004,takenaka2005,jonsson2007,kaiser2010,jaramillo2014,biswas2020,pustogow2021,uzkur2021,uykur2022}. The common feature in all of these experimental findings is that the displaced Drude peak position is an increasing function of temperature, $\omega_{MAX} \sim T^\alpha$, with the coefficient $\alpha$ falling in the range, $0<\alpha<3/2$ \cite{fratini2021}. To at least qualitatively compare our results with this robust experimental feature, we show in Fig.~\ref{fig:max_temp_dependence} the temperature dependence of the position of the maxima in both the current-current correlation function and optical conductivity, as well as the temperature dependence of the maximum values themselves. For the parameter set $S_{\text{2D}}$, we focus only on the $\pi$-ton vertex contribution since the displaced Drude peak is not yet present in the total optical conductivity, while for the  parameter set $S_{\text{1D}}$ we consider the total, bubble plus $\pi$-ton, contribution as well.
	
	In both the 1D and 2D systems, the position of the maxima of the $\pi$-ton vertex contributions, $\chi_{VERT}$ and $\sigma_{VERT}$ in Figs.~\ref{fig:max_temp_dependence}(b,d), is clearly an increasing function of temperature, with a roughly linear dependence of the maximum position, $\omega_{MAX} \sim (T-T_c)$. When it comes to the total optical conductivity in the 1D case, on the other hand, the displaced Drude peak position within the current approximations is rather a decreasing function of temperature. This behavior is not in line with the aforementioned experimental results. However, one can argue that it might also significantly depend on the fine balance between the spectral weight of the bubble and the $\pi$-ton vertex contributions. 
	
	Finally, when it comes to the maximum values of $\chi_{VERT}$ and $\sigma_{VERT}$ and the height of the displaced Drude peak in the total optical conductivity, from Figs.~\ref{fig:max_temp_dependence}(a,c) it is evident that they all increase as the phase transition is approached. Interestingly, in the 1D case we roughly have $\underset{\omega>0}{\max}\;\sigma_{VERT}(\omega) \sim \left( T-T_c\right)^{-1} $, while in the 2D case $\underset{\omega>0}{\max}\;\sigma_{VERT}(\omega)  \sim \ln\left[ \left( T-T_c\right)^{-1} \right] $ (note the logarithmic scale). This scaling further supports the formation of a displaced Drude peak as a result of the $\pi$-ton vertex corrections in 2D systems, with the caveat that any firm conclusions are difficult to draw due to logarithmic scaling.
	
 \subsection{Comments on the relation to previous results}
	\label{ssec:pastresults}
	The displaced Drude peak was not observed in the earlier RPA treatments of the $\pi$-ton vertex corrections \cite{worm2021,simard_eq,simard_noneq}. To  understand the discrepancies between those previous and the present results, we here recalculate the bubble and the $\pi$-ton vertex contributions to the current-current correlation function on the imaginary axis for the parameter set $S_{\text{DMFT}}$ and the temperature $T = 0.08$ following the approach of Ref.~\cite{simard_eq}. 
 
\begin{figure*}
		\centering
		\includegraphics[width=.99\textwidth]{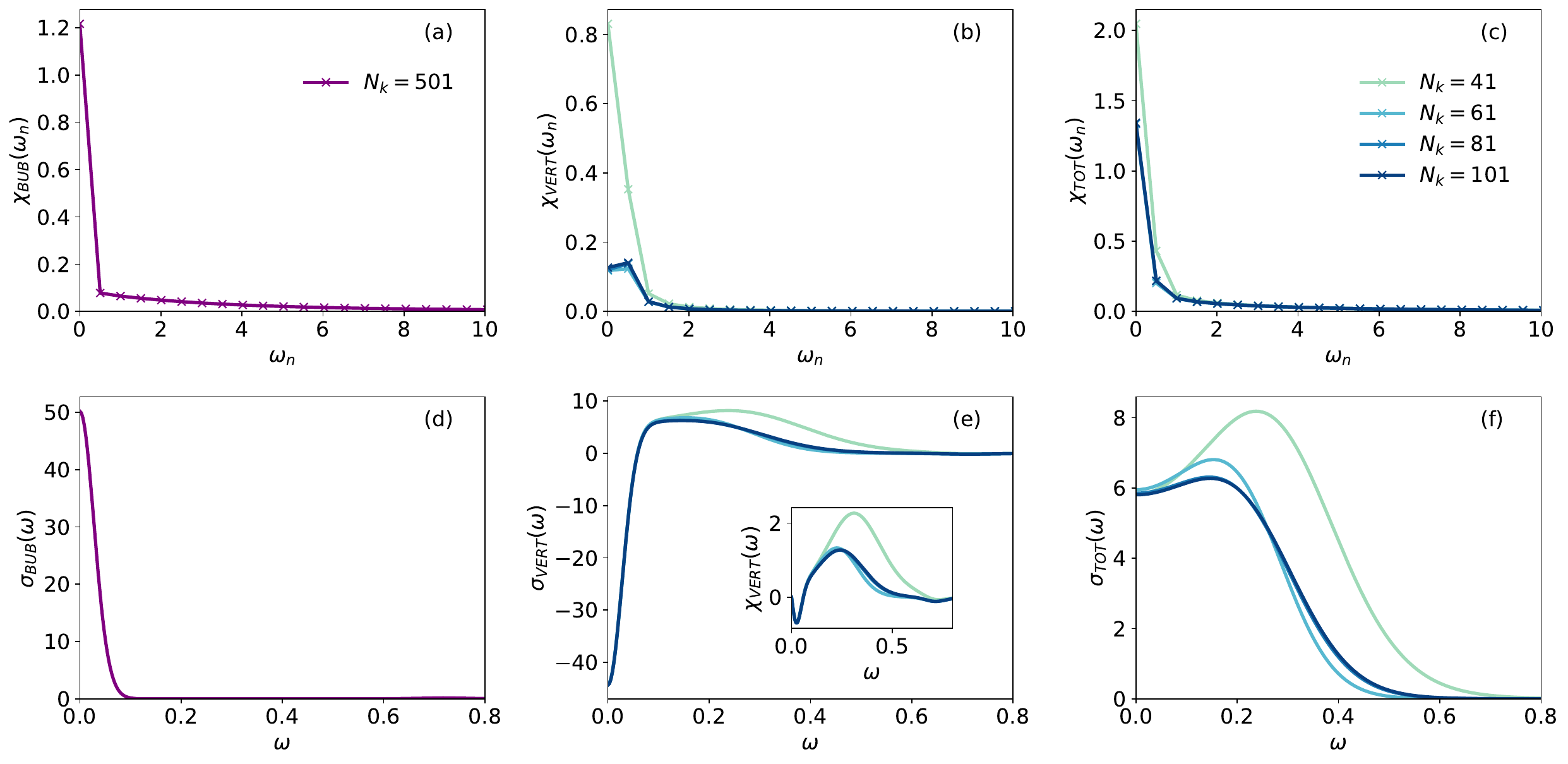}
		\renewcommand{\figurename}{Fig.}
		\caption{(a-c) Bubble, RPA $\pi$-ton vertex, and total contribution to the current-current correlation function on the Matsubara axis calculated for the parameter set $S_{\text{DMFT}}$ with $T=0.08$ and several momentum grids $N_k$. (d-f) Corresponding optical conductivity spectra on the real-frequency axis. The inset of panel (e) further shows the current-current correlation function for real frequencies.}
		\label{fig:optical_conductivity_matsubara_RPA}
	\end{figure*}

 Motivated by the convergence challenges discussed in Section~\ref{ssec:fate_vertex}, in Fig.~\ref{fig:optical_conductivity_matsubara_RPA}(b) we show the $\pi$-ton vertex contributions on the Matsubara axis for four different momentum grids. It is apparent that $\chi_{VERT}$ for the grid $N_k = 41$, which roughly equals the number of momentum points used previously in Refs.~\cite{simard_eq,simard_noneq}, differs quite significantly from the results for denser momentum grids. Not only is the magnitude of $\chi_{VERT}$  roughly four times smaller for the first few Matsubara frequencies but more importantly, the slope between zero and the first Matsubara frequency becomes positive in the latter cases. Those differences can be even more pronounced for temperatures closer to the transition boundary, such as those considered in Ref.~\cite{simard_eq}.
	
	In order to obtain the optical conductivity spectra on the real-frequency axis from the current-current correlation functions in Figs.~\ref{fig:optical_conductivity_matsubara_RPA}(a-c), we carry out an analytic continuation using the ana\_cont package \cite{ana_cont}. We analytically continue only $\chi_{BUB}$ and $\chi_{TOT}$, while the $\pi$-ton vertex contributions are obtained as the difference $\sigma_{VERT}=\sigma_{TOT}-\sigma_{BUB}$. More details on the analytic continuation are given in  Appendix~\ref{app:ana_cont}. 
	
	In Fig.~\ref{fig:optical_conductivity_matsubara_RPA}(d-f), we show the resulting optical conductivity spectra for an optimal set of analytic continuation parameters. We conclude that the overall shape of the $\pi$-ton vertex corrections $\sigma_{VERT}$ is quite robust to different analytic continuation parameters, that $\sigma_{VERT}$ is converged for $N_k = 81$, and that it actually shows a similar qualitative behavior as the shape of the $\pi$-ton vertex corrections predicted  in Sections~\ref{subsec:real_axis} and~\ref{ssec:fate_vertex}. Namely, the $\pi$-ton vertex corrections, Fig.~\ref{fig:optical_conductivity_matsubara_RPA}(e), tend to suppress the optical conductivity at low frequencies and enhance it at higher ones, producing together with the bubble contribution the displaced Drude peak, see Fig.~\ref{fig:optical_conductivity_matsubara_RPA}(f). 
	
	The enhancement of the current-current correlation function by the $\pi$-ton vertex contributions for the case with $N_k=41$, see inset of  Fig.~\ref{fig:optical_conductivity_matsubara_RPA}(e), is maximal around $\omega\approx0.31$, while for low frequencies, $\omega < 0.06$, the $\pi$-ton vertex contributions are rather negative. For exactly the same parameters, the $\pi$-ton vertex contributions were calculated directly on the real-frequency axis in Ref.~\cite{simard_noneq}. While a broad peak, similar to the one in the inset of  Fig.~\ref{fig:optical_conductivity_matsubara_RPA}(e), was reported around $\omega\approx0.35$, the $\pi$-ton vertex contributions were reported to be positive all the way down to the zero frequency. We note, however, that in Ref.~\cite{simard_noneq} a Fourier transformation with  time window $\Delta t = 17$ was used, which implies a smearing of the low-frequency features, $\omega < \left( \Delta t\right) ^{-1}$. This might explain why no negative vertex correction was observed at small frequencies.
	
	As a last remark, we note that the results in Fig.~\ref{fig:optical_conductivity_matsubara_RPA} are obtained with DMFT Green's functions which, apart from the quasiparticle pole, contain the incoherent contribution corresponding to the Hubbard bands. Our results suggest that for relatively weak interactions the Hubbard bands play little to no role in shaping the qualitative behavior of the $\pi$-ton vertex corrections. In other words, the displaced Drude peak is obtained solely by considering vertex corrections of the low-energy  quasiparticle excitations. It is not an additional peak but as if the Drude peak itself was shifted away from $\omega = 0$.
	
	\section{Conclusions}
	\label{sec:conclusion}
	
	Our main finding is that $\pi$-vertex contributions result in a displaced Drude peak for correlated metals, a true shift of the Drude peak to a maximum at a nonzero frequency.
	The presence of a displaced Drude peak in the 1D case is unambiguous with a linear scaling of the position and maximum of the displaced Drude peak with temperature (upon approaching the antiferromagnetic phase-transition temperature $T_N$).
	In the 2D case, we observe qualitatively the same features, however, the logarithmic scaling in $T-T_N$  (instead of linear in 1D) makes unambiguous calculations practically impossible.

	The displaced Drude peak  was confirmed using the same two methods employed in earlier RPA studies of the $\pi$-ton contribution. That is the direct calculation for real frequencies and the analytic continuation of the Matsubara frequency calculation. As before, we employ a constant self-energy for the former  and  the IPT self-energy for the latter. 
	Our paper is thus able to  reconcile the apparent discrepancies between earlier results on $\pi$-ton vertex corrections in weakly correlated metals. 
	These reported a broadening and sharpening of the Drude peak \cite{worm2021}, and an additional peak in the optical conductivity \cite{simard_noneq,simard_eq}.
	Let us emphasize the importance of a proper $\nu$- and $\k$-grid convergence. This is challenging to achieve close to the antiferromagnetic phase transition, where $\pi$-ton effects become more pronounced and lead to the displaced Drude peak.

	Our work shows that besides disorder localization \cite{abrahams1979,gorkov1979,gotze1979,altshuler1980} and external phonon modes close to zero frequency \cite{fratini2021,rammal2023}, 
	displaced Drude peaks should also be expected in general close to antiferromagnetic and charge density wave transitions. Our work hence shows an additional route to the displaced Drude peak: $\pi$-ton vertex corrections due to strong antiferromagnetic and charge density wave fluctuations. The effect is expected to be larger in 1D than in 2D. 
	
	One remaining question is: How can we distinguish this 
	$\pi$-ton physics in the transversal particle-hole channel from localization effects in the particle-particle channel or external bosons such as low energy phonons? One strategy for unambiguously identifying $\pi$-tons in an experiment would be to look at the change of the displaced Drude peak when approaching, e.g., an antiferromagnetic phase transition. We predict that if it is a $\pi$-ton peak, the overall effect will be enhanced closer to the phase transition. 
	
	\begin{acknowledgments}
		We thank P. Worm for useful discussions. The authors  acknowledge the support of the Research Unit QUAST by the Deutsche Foschungsgemeinschaft  (DFG; project ID FOR5249), the Austrian Science Fund (FWF; project ID I 5868), and the Swiss National Science Foundation (SNSF). J.K., A.K., and K.H. also acknowledge support of the FWF Projects No. P 36213 and V1018. 

The data of this paper is available at XXX.

	\end{acknowledgments}

	\begin{widetext}
		\appendix
		
		\section{Fit of the Ornstein-Zernike form to the RPA $\pi$-ton vertex function} \label{app:oz_fit}
		
		We extract the Ornstein-Zernike parameters $A$, $\lambda$, and $\xi$ in Eq.~\eqref{eq:oz_form} from the RPA $\pi$-ton vertex function in Eq.~\eqref{eq:RPA_vertex} in three steps. First, we calculate the real part of the static RPA $\pi$-ton vertex function around the wave vector $q=\pi$, as well as the low-frequency limit of the RPA $\pi$-ton vertex function at $q=\pi$, using dense momentum and frequency grids, with $N_k = 1000$ and $N_\nu = 7993$ points, respectively. Then in the second step, we consider the inverse of the static Ornstein-Zernike function around $q = \pi$
		
		\begin{equation}
			\left[ F^{OZ}_{q\approx \pi,\omega}\right]^{-1} \approx \frac{1}{A}q^2 + \frac{\xi^{-2}}{A}\;,
		\end{equation}
		and fit the second order polynomial, $aq^2 + c$, to the calculated static RPA $\pi$-ton  vertex function around $q\approx \pi$. The inverse of the coefficient $a$ gives us the parameter $A=a^{-1}$, while the correlation length is obtained as $\xi = \sqrt{a/c}$.
		
		In the last (third) step we fit the frequency dependence. To this end, we consider the inverse of the real part of the Ornstein-Zernike function at $q = \pi$
		
		\begin{equation}
			\left[ F^{OZ}_{q=\pi,\omega}\right]^{-1} \approx \frac{\lambda^2}{A\xi^{-2}}\omega^2 + \frac{A}{\xi^{-4}}\;,
		\end{equation}
		and again fit a second order polynomial, $a'\omega^2 +c'$, now to the low frequency part of the RPA $\pi$-ton vertex function at $q=\pi$. We use the coefficient $a'$ together with the previously extracted $A$ and $\xi$ to finally obtain $\lambda = \sqrt{a'A}/\xi$.
		
		Following the above procedure, we extract from the RPA $\pi$-ton vertex function the Ornstein-Zernike parameters for the parameter set $S_{\text{1D}}$ and inverse temperatures $\beta = 16-22$. The corresponding parameters $A$, $\lambda$, and $\xi$ are shown in Table~\ref{table:oz-parameters}, while in Fig.~\ref{fig:oz_fit}, we show the RPA $\pi$-ton vertex function together with its approximate Ornstein-Zernike form for the inverse temperature $\beta = 17$.
		
		\begin{figure*}
			\centering
			\includegraphics[width=1\textwidth]{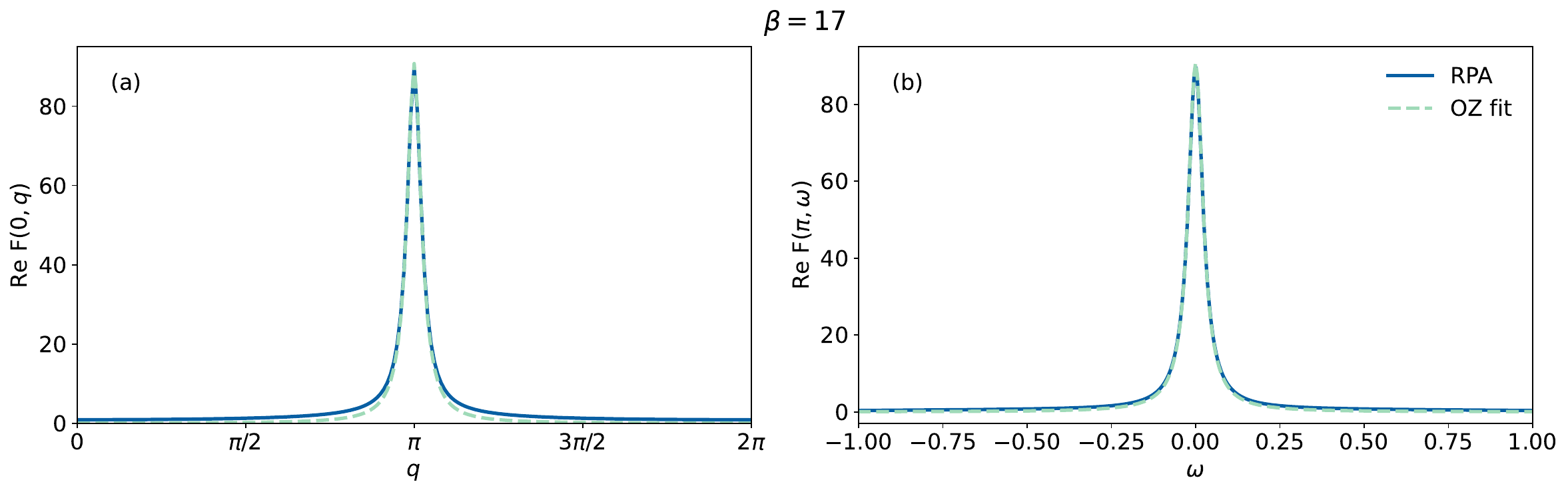}
			\renewcommand{\figurename}{Fig.}
			\caption{Real part of the Ornstein-Zernike vertex function (teal dashed line) fitted to the RPA $\pi$-ton vertex function (blue solid line) for the parameter set $S_{\text{1D}}$ and the inverse temperature $\beta = 17$, as a function of (a) $q$ for a fixed $\omega=0$, and (b) $\omega$ for a  fixed $q=\pi$.}
			\label{fig:oz_fit}
		\end{figure*}
		
		\section{Details on the analytic continuation of the current-current correlation function} \label{app:ana_cont}
		
		The analytic continuation of the current-current correlation functions $\chi_{jj}(i\omega_n)$ in Figs.~\ref{fig:optical_conductivity_matsubara_RPA}(a)-(c) is carried out by means of the maximum entropy method as implemented within the ana\_cont package \cite{ana_cont}. The solver of ana\_cont directly returns $\sigma(\omega)$, which is related to  $\chi_{jj}(i\omega_n)$ with a kernel $K_b(\omega_n,\omega)=\frac{\omega^2}{\omega^2_n+\omega^2}$ as $\chi_{jj}(i\omega_n) = \int_{0}^{\infty}d\omega K_b(\omega_n,\omega) \sigma(\omega)$. We analytically continue only the bubble and the total contributions, while the $\pi$-ton vertex contributions to the optical conductivity are obtained as the difference, $\sigma_{VERT}=\sigma_{TOT}-\sigma_{BUB}$. As input to our analytic continuation problem, we use the values $\chi_{jj}(i\omega_n)$ for the first 30 positive Matsubara frequencies, including the zero frequency, normalized with respect to the value of $\chi_{jj}$ at zero Matsubara frequency. Additionally, we set the amplitude of error to $5\cdot 10^5$ and $5\cdot 10^3$ for the bubble and the total contribution, respectively. The corresponding real frequency spectra are computed on the grid $[0, \lfloor \frac{2\pi }{\beta}\cdot30\rfloor]$ with 5000 points. 
		
		The method for determining the hyperparameter $\alpha$ and the model for the prior probability of the spectrum have a significant influence on the behavior of the resulting spectra; see Ref.~\cite{ana_cont} for further information on these parameters, the chi2kink method and Gaussian broadening. For example, in Fig.~\eqref{fig:ana_cont} we show the spectra for the parameter set $S_{\text{DMFT}}$, calculated with $N_k=101$ momentum points, and two inverse temperatures $\beta = 12$ and $\beta = 17$, obtained with the chi2kink method and employing several models. In particular, we compare the spectra obtained with the flat model and Gaussian functions with various widths $\sigma$. Although the displaced Drude peak is clearly visible in all cases considered, choosing a thinner Gaussian leads to a more pronounced  displaced Drude peak behavior. This is particularly evident in the case with $\sigma = 1$, where the $\pi$-ton vertex corrections may appear to tend to peak around $\omega\sim 0.2$ as the temperature is lowered, see Figs.~\ref{fig:ana_cont}(b) and (e).
		
		In the main text, we present the results obtained with the chi2kink method and the flat model, in order to avoid any bias towards an extremely sharp displaced Drude peak in the total optical conductivity. Although, as expected, the displaced Drude peak is more pronounced for the inverse temperature $\beta=17$, in the main text we present the case with $\beta=12.5$ to enable a direct comparison of our new results with those obtained in Ref.~\cite{simard_noneq}.
		
		\begin{figure*}
			\centering
			\includegraphics[width=1\textwidth]{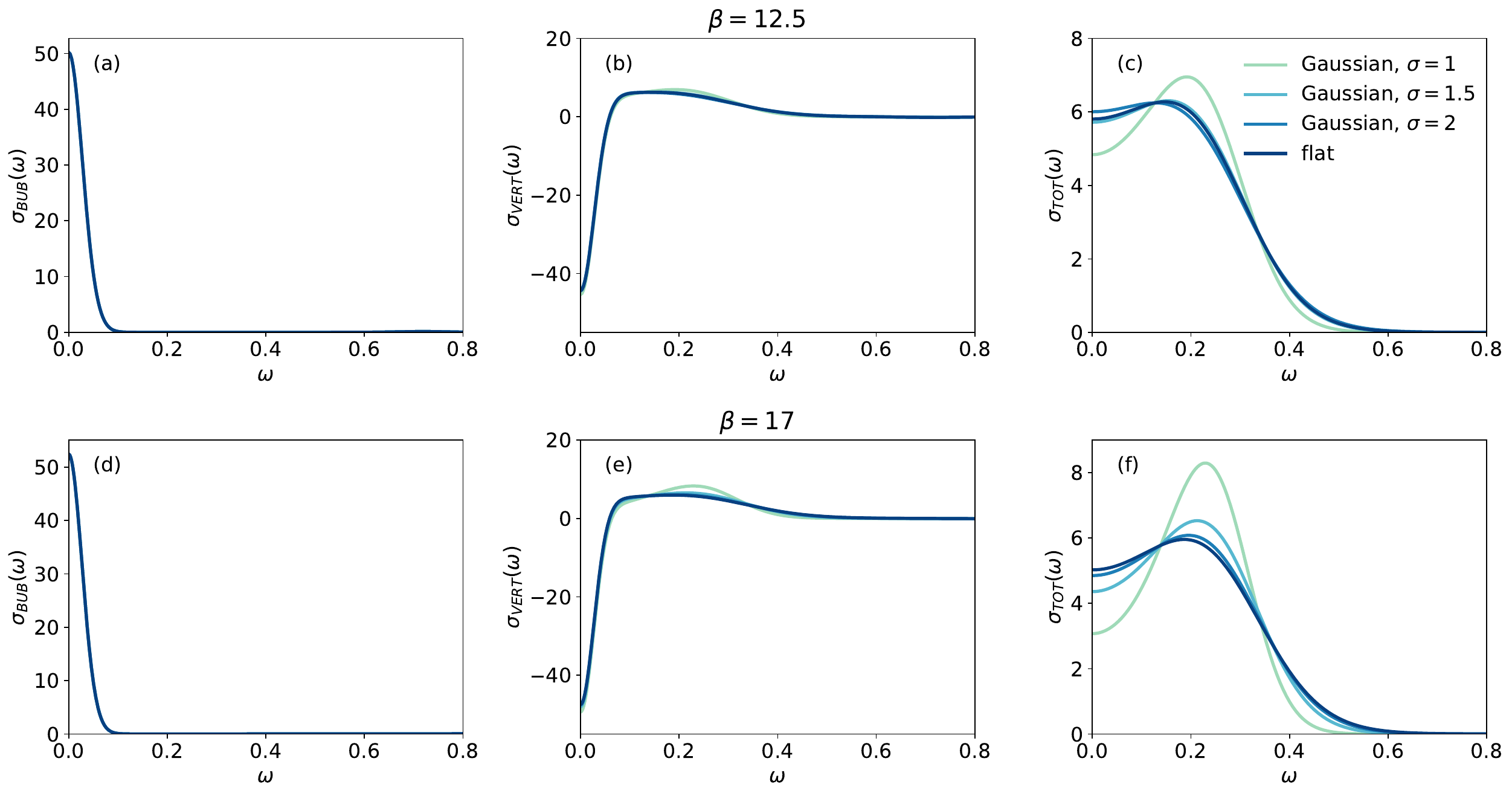}
			\renewcommand{\figurename}{Fig.}
			\caption{Bubble (a,d), $\pi$-ton vertex (b,e), and total contribution (c,f) to the optical conductivity obtained with the analytic continuation using the chi2kink method and several different default models, for the parameter set $S_{\text{DMFT}}$, the momentum grid $N_k =101$, and two temperatures $\beta=12.5$ (a-c) and $\beta=17$ (d-f). }
			\label{fig:ana_cont}
		\end{figure*}
		
	\end{widetext}

	\bibliography{paper}

\begin{thebibliography}{71}%
\makeatletter
\providecommand \@ifxundefined [1]{%
 \@ifx{#1\undefined}
}%
\providecommand \@ifnum [1]{%
 \ifnum #1\expandafter \@firstoftwo
 \else \expandafter \@secondoftwo
 \fi
}%
\providecommand \@ifx [1]{%
 \ifx #1\expandafter \@firstoftwo
 \else \expandafter \@secondoftwo
 \fi
}%
\providecommand \natexlab [1]{#1}%
\providecommand \enquote  [1]{``#1''}%
\providecommand \bibnamefont  [1]{#1}%
\providecommand \bibfnamefont [1]{#1}%
\providecommand \citenamefont [1]{#1}%
\providecommand \href@noop [0]{\@secondoftwo}%
\providecommand \href [0]{\begingroup \@sanitize@url \@href}%
\providecommand \@href[1]{\@@startlink{#1}\@@href}%
\providecommand \@@href[1]{\endgroup#1\@@endlink}%
\providecommand \@sanitize@url [0]{\catcode `\\12\catcode `\$12\catcode
  `\&12\catcode `\#12\catcode `\^12\catcode `\_12\catcode `\%12\relax}%
\providecommand \@@startlink[1]{}%
\providecommand \@@endlink[0]{}%
\providecommand \url  [0]{\begingroup\@sanitize@url \@url }%
\providecommand \@url [1]{\endgroup\@href {#1}{\urlprefix }}%
\providecommand \urlprefix  [0]{URL }%
\providecommand \Eprint [0]{\href }%
\providecommand \doibase [0]{https://doi.org/}%
\providecommand \selectlanguage [0]{\@gobble}%
\providecommand \bibinfo  [0]{\@secondoftwo}%
\providecommand \bibfield  [0]{\@secondoftwo}%
\providecommand \translation [1]{[#1]}%
\providecommand \BibitemOpen [0]{}%
\providecommand \bibitemStop [0]{}%
\providecommand \bibitemNoStop [0]{.\EOS\space}%
\providecommand \EOS [0]{\spacefactor3000\relax}%
\providecommand \BibitemShut  [1]{\csname bibitem#1\endcsname}%
\let\auto@bib@innerbib\@empty
\bibitem [{\citenamefont {Drude}(1900{\natexlab{a}})}]{drude1}%
  \BibitemOpen
  \bibfield  {author} {\bibinfo {author} {\bibfnamefont {P.}~\bibnamefont
  {Drude}},\ }\bibfield  {title} {\bibinfo {title} {{Zur Elektronentheorie der
  Metalle}},\ }\href {https://doi.org/https://doi.org/10.1002/andp.19003060312}
  {\bibfield  {journal} {\bibinfo  {journal} {Annalen der Physik}\ }\textbf
  {\bibinfo {volume} {306}},\ \bibinfo {pages} {566} (\bibinfo {year}
  {1900}{\natexlab{a}})}\BibitemShut {NoStop}%
\bibitem [{\citenamefont {Drude}(1900{\natexlab{b}})}]{drude2}%
  \BibitemOpen
  \bibfield  {author} {\bibinfo {author} {\bibfnamefont {P.}~\bibnamefont
  {Drude}},\ }\bibfield  {title} {\bibinfo {title} {{Zur Elektronentheorie der
  Metalle; II. Teil. Galvanomagnetische und thermomagnetische Effecte}},\
  }\href {https://doi.org/https://doi.org/10.1002/andp.19003081102} {\bibfield
  {journal} {\bibinfo  {journal} {Annalen der Physik}\ }\textbf {\bibinfo
  {volume} {308}},\ \bibinfo {pages} {369} (\bibinfo {year}
  {1900}{\natexlab{b}})}\BibitemShut {NoStop}%
\bibitem [{\citenamefont {Frenkel}(1931)}]{frenkel1931}%
  \BibitemOpen
  \bibfield  {author} {\bibinfo {author} {\bibfnamefont {J.}~\bibnamefont
  {Frenkel}},\ }\bibfield  {title} {\bibinfo {title} {{On the Transformation of
  light into Heat in Solids. I}},\ }\href
  {https://doi.org/10.1103/PhysRev.37.17} {\bibfield  {journal} {\bibinfo
  {journal} {Phys. Rev.}\ }\textbf {\bibinfo {volume} {37}},\ \bibinfo {pages}
  {17} (\bibinfo {year} {1931})}\BibitemShut {NoStop}%
\bibitem [{\citenamefont {Wannier}(1937)}]{wannier1937}%
  \BibitemOpen
  \bibfield  {author} {\bibinfo {author} {\bibfnamefont {G.~H.}\ \bibnamefont
  {Wannier}},\ }\bibfield  {title} {\bibinfo {title} {{The Structure of
  Electronic Excitation Levels in Insulating Crystals}},\ }\href
  {https://doi.org/10.1103/PhysRev.52.191} {\bibfield  {journal} {\bibinfo
  {journal} {Phys. Rev.}\ }\textbf {\bibinfo {volume} {52}},\ \bibinfo {pages}
  {191} (\bibinfo {year} {1937})}\BibitemShut {NoStop}%
\bibitem [{\citenamefont {Abrahams}\ \emph {et~al.}(1979)\citenamefont
  {Abrahams}, \citenamefont {Anderson}, \citenamefont {Licciardello},\ and\
  \citenamefont {Ramakrishnan}}]{abrahams1979}%
  \BibitemOpen
  \bibfield  {author} {\bibinfo {author} {\bibfnamefont {E.}~\bibnamefont
  {Abrahams}}, \bibinfo {author} {\bibfnamefont {P.~W.}\ \bibnamefont
  {Anderson}}, \bibinfo {author} {\bibfnamefont {D.~C.}\ \bibnamefont
  {Licciardello}},\ and\ \bibinfo {author} {\bibfnamefont {T.~V.}\ \bibnamefont
  {Ramakrishnan}},\ }\bibfield  {title} {\bibinfo {title} {{Scaling Theory of
  Localization: Absence of Quantum Diffusion in Two Dimensions}},\ }\href
  {https://doi.org/10.1103/PhysRevLett.42.673} {\bibfield  {journal} {\bibinfo
  {journal} {Phys. Rev. Lett.}\ }\textbf {\bibinfo {volume} {42}},\ \bibinfo
  {pages} {673} (\bibinfo {year} {1979})}\BibitemShut {NoStop}%
\bibitem [{\citenamefont {{Gor'kov}}\ \emph {et~al.}(1996)\citenamefont
  {{Gor'kov}}, \citenamefont {{Larkin}},\ and\ \citenamefont
  {{Khmel'Nitskii}}}]{gorkov1979}%
  \BibitemOpen
  \bibfield  {author} {\bibinfo {author} {\bibfnamefont {L.~P.}\ \bibnamefont
  {{Gor'kov}}}, \bibinfo {author} {\bibfnamefont {A.~I.}\ \bibnamefont
  {{Larkin}}},\ and\ \bibinfo {author} {\bibfnamefont {D.~E.}\ \bibnamefont
  {{Khmel'Nitskii}}},\ }\bibfield  {title} {\bibinfo {title} {{{Particle
  conductivity in a two-dimensional random potential}}},\ }in\ \href
  {https://doi.org/10.1142/9789814317344_0022} {\emph {\bibinfo {booktitle} {30
  Years of the Landau Institute - Selected Papers. Edited by KHALATNIKOV ISAAK
  M ET AL. Published by World Scientific Publishing Co. Pte. Ltd}}}\ (\bibinfo
  {year} {1996})\ pp.\ \bibinfo {pages} {157--161}\BibitemShut {NoStop}%
\bibitem [{\citenamefont {Götze}\ \emph {et~al.}(1979)\citenamefont {Götze},
  \citenamefont {Prelovšek},\ and\ \citenamefont {Wölfle}}]{gotze1979}%
  \BibitemOpen
  \bibfield  {author} {\bibinfo {author} {\bibfnamefont {W.}~\bibnamefont
  {Götze}}, \bibinfo {author} {\bibfnamefont {P.}~\bibnamefont {Prelovšek}},\
  and\ \bibinfo {author} {\bibfnamefont {P.}~\bibnamefont {Wölfle}},\
  }\bibfield  {title} {\bibinfo {title} {{Localization of particles in a
  two-dimensional random potential}},\ }\href
  {https://doi.org/https://doi.org/10.1016/0038-1098(79)90654-9} {\bibfield
  {journal} {\bibinfo  {journal} {Solid State Communications}\ }\textbf
  {\bibinfo {volume} {30}},\ \bibinfo {pages} {369} (\bibinfo {year}
  {1979})}\BibitemShut {NoStop}%
\bibitem [{\citenamefont {Altshuler}\ \emph {et~al.}(1980)\citenamefont
  {Altshuler}, \citenamefont {Khmel'nitzkii}, \citenamefont {Larkin},\ and\
  \citenamefont {Lee}}]{altshuler1980}%
  \BibitemOpen
  \bibfield  {author} {\bibinfo {author} {\bibfnamefont {B.~L.}\ \bibnamefont
  {Altshuler}}, \bibinfo {author} {\bibfnamefont {D.}~\bibnamefont
  {Khmel'nitzkii}}, \bibinfo {author} {\bibfnamefont {A.~I.}\ \bibnamefont
  {Larkin}},\ and\ \bibinfo {author} {\bibfnamefont {P.~A.}\ \bibnamefont
  {Lee}},\ }\bibfield  {title} {\bibinfo {title} {{Magnetoresistance and Hall
  effect in a disordered two-dimensional electron gas}},\ }\href
  {https://doi.org/10.1103/PhysRevB.22.5142} {\bibfield  {journal} {\bibinfo
  {journal} {Phys. Rev. B}\ }\textbf {\bibinfo {volume} {22}},\ \bibinfo
  {pages} {5142} (\bibinfo {year} {1980})}\BibitemShut {NoStop}%
\bibitem [{\citenamefont {Clarke}(1993)}]{clarke1993}%
  \BibitemOpen
  \bibfield  {author} {\bibinfo {author} {\bibfnamefont {D.~G.}\ \bibnamefont
  {Clarke}},\ }\bibfield  {title} {\bibinfo {title} {{Particle-hole bound
  states in Mott-Hubbard insulators}},\ }\href
  {https://doi.org/10.1103/PhysRevB.48.7520} {\bibfield  {journal} {\bibinfo
  {journal} {Phys. Rev. B}\ }\textbf {\bibinfo {volume} {48}},\ \bibinfo
  {pages} {7520} (\bibinfo {year} {1993})}\BibitemShut {NoStop}%
\bibitem [{\citenamefont {Maebashi}\ and\ \citenamefont
  {Fukuyama}(1997)}]{maebashi1997}%
  \BibitemOpen
  \bibfield  {author} {\bibinfo {author} {\bibfnamefont {H.}~\bibnamefont
  {Maebashi}}\ and\ \bibinfo {author} {\bibfnamefont {H.}~\bibnamefont
  {Fukuyama}},\ }\bibfield  {title} {\bibinfo {title} {{Electrical Conductivity
  of Interacting Fermions. I. General Formulation}},\ }\href
  {https://doi.org/10.1143/JPSJ.66.3577} {\bibfield  {journal} {\bibinfo
  {journal} {Journal of the Physical Society of Japan}\ }\textbf {\bibinfo
  {volume} {66}},\ \bibinfo {pages} {3577} (\bibinfo {year}
  {1997})}\BibitemShut {NoStop}%
\bibitem [{\citenamefont {Essler}\ \emph {et~al.}(2001)\citenamefont {Essler},
  \citenamefont {Gebhard},\ and\ \citenamefont {Jeckelmann}}]{essler2001}%
  \BibitemOpen
  \bibfield  {author} {\bibinfo {author} {\bibfnamefont {F.~H.~L.}\
  \bibnamefont {Essler}}, \bibinfo {author} {\bibfnamefont {F.}~\bibnamefont
  {Gebhard}},\ and\ \bibinfo {author} {\bibfnamefont {E.}~\bibnamefont
  {Jeckelmann}},\ }\bibfield  {title} {\bibinfo {title} {{Excitons in
  one-dimensional Mott insulators}},\ }\href
  {https://doi.org/10.1103/PhysRevB.64.125119} {\bibfield  {journal} {\bibinfo
  {journal} {Phys. Rev. B}\ }\textbf {\bibinfo {volume} {64}},\ \bibinfo
  {pages} {125119} (\bibinfo {year} {2001})}\BibitemShut {NoStop}%
\bibitem [{\citenamefont {Wr\'obel}\ and\ \citenamefont
  {Eder}(2002)}]{wrobel2002}%
  \BibitemOpen
  \bibfield  {author} {\bibinfo {author} {\bibfnamefont {P.}~\bibnamefont
  {Wr\'obel}}\ and\ \bibinfo {author} {\bibfnamefont {R.}~\bibnamefont
  {Eder}},\ }\bibfield  {title} {\bibinfo {title} {{Excitons in Mott
  insulators}},\ }\href {https://doi.org/10.1103/PhysRevB.66.035111} {\bibfield
   {journal} {\bibinfo  {journal} {Phys. Rev. B}\ }\textbf {\bibinfo {volume}
  {66}},\ \bibinfo {pages} {035111} (\bibinfo {year} {2002})}\BibitemShut
  {NoStop}%
\bibitem [{\citenamefont {Jeckelmann}(2003)}]{jeckelmann2003}%
  \BibitemOpen
  \bibfield  {author} {\bibinfo {author} {\bibfnamefont {E.}~\bibnamefont
  {Jeckelmann}},\ }\bibfield  {title} {\bibinfo {title} {{Optical excitations
  in a one-dimensional Mott insulator}},\ }\href
  {https://doi.org/10.1103/PhysRevB.67.075106} {\bibfield  {journal} {\bibinfo
  {journal} {Phys. Rev. B}\ }\textbf {\bibinfo {volume} {67}},\ \bibinfo
  {pages} {075106} (\bibinfo {year} {2003})}\BibitemShut {NoStop}%
\bibitem [{\citenamefont {Kontani}(2006)}]{kontani2006}%
  \BibitemOpen
  \bibfield  {author} {\bibinfo {author} {\bibfnamefont {H.}~\bibnamefont
  {Kontani}},\ }\bibfield  {title} {\bibinfo {title} {{Optical Conductivity and
  Hall Coefficient in High-Tc Superconductors: Significant Role of Current
  Vertex Corrections}},\ }\href {https://doi.org/10.1143/JPSJ.75.013703}
  {\bibfield  {journal} {\bibinfo  {journal} {Journal of the Physical Society
  of Japan}\ }\textbf {\bibinfo {volume} {75}},\ \bibinfo {pages} {013703}
  (\bibinfo {year} {2006})}\BibitemShut {NoStop}%
\bibitem [{\citenamefont {Lin}\ \emph {et~al.}(2009)\citenamefont {Lin},
  \citenamefont {Gull},\ and\ \citenamefont {Millis}}]{lin2009}%
  \BibitemOpen
  \bibfield  {author} {\bibinfo {author} {\bibfnamefont {N.}~\bibnamefont
  {Lin}}, \bibinfo {author} {\bibfnamefont {E.}~\bibnamefont {Gull}},\ and\
  \bibinfo {author} {\bibfnamefont {A.~J.}\ \bibnamefont {Millis}},\ }\bibfield
   {title} {\bibinfo {title} {{Optical conductivity from cluster dynamical
  mean-field theory: Formalism and application to high-temperature
  superconductors}},\ }\href {https://doi.org/10.1103/PhysRevB.80.161105}
  {\bibfield  {journal} {\bibinfo  {journal} {Phys. Rev. B}\ }\textbf {\bibinfo
  {volume} {80}},\ \bibinfo {pages} {161105} (\bibinfo {year}
  {2009})}\BibitemShut {NoStop}%
\bibitem [{\citenamefont {Bergeron}\ \emph {et~al.}(2011)\citenamefont
  {Bergeron}, \citenamefont {Hankevych}, \citenamefont {Kyung},\ and\
  \citenamefont {Tremblay}}]{bergeron2011}%
  \BibitemOpen
  \bibfield  {author} {\bibinfo {author} {\bibfnamefont {D.}~\bibnamefont
  {Bergeron}}, \bibinfo {author} {\bibfnamefont {V.}~\bibnamefont {Hankevych}},
  \bibinfo {author} {\bibfnamefont {B.}~\bibnamefont {Kyung}},\ and\ \bibinfo
  {author} {\bibfnamefont {A.-M.~S.}\ \bibnamefont {Tremblay}},\ }\bibfield
  {title} {\bibinfo {title} {{Optical and dc conductivity of the
  two-dimensional Hubbard model in the pseudogap regime and across the
  antiferromagnetic quantum critical point including vertex corrections}},\
  }\href {https://doi.org/10.1103/PhysRevB.84.085128} {\bibfield  {journal}
  {\bibinfo  {journal} {Phys. Rev. B}\ }\textbf {\bibinfo {volume} {84}},\
  \bibinfo {pages} {085128} (\bibinfo {year} {2011})}\BibitemShut {NoStop}%
\bibitem [{\citenamefont {Chubukov}\ \emph {et~al.}(2014)\citenamefont
  {Chubukov}, \citenamefont {Maslov},\ and\ \citenamefont
  {Yudson}}]{chubukov2014}%
  \BibitemOpen
  \bibfield  {author} {\bibinfo {author} {\bibfnamefont {A.~V.}\ \bibnamefont
  {Chubukov}}, \bibinfo {author} {\bibfnamefont {D.~L.}\ \bibnamefont
  {Maslov}},\ and\ \bibinfo {author} {\bibfnamefont {V.~I.}\ \bibnamefont
  {Yudson}},\ }\bibfield  {title} {\bibinfo {title} {{Optical conductivity of a
  two-dimensional metal at the onset of spin-density-wave order}},\ }\href
  {https://doi.org/10.1103/PhysRevB.89.155126} {\bibfield  {journal} {\bibinfo
  {journal} {Phys. Rev. B}\ }\textbf {\bibinfo {volume} {89}},\ \bibinfo
  {pages} {155126} (\bibinfo {year} {2014})}\BibitemShut {NoStop}%
\bibitem [{\citenamefont {Kokalj}(2017)}]{kokalj2017}%
  \BibitemOpen
  \bibfield  {author} {\bibinfo {author} {\bibfnamefont {J.}~\bibnamefont
  {Kokalj}},\ }\bibfield  {title} {\bibinfo {title} {{Bad-metallic behavior of
  doped Mott insulators}},\ }\href {https://doi.org/10.1103/PhysRevB.95.041110}
  {\bibfield  {journal} {\bibinfo  {journal} {Phys. Rev. B}\ }\textbf {\bibinfo
  {volume} {95}},\ \bibinfo {pages} {041110} (\bibinfo {year}
  {2017})}\BibitemShut {NoStop}%
\bibitem [{\citenamefont {Maslov}\ and\ \citenamefont
  {Chubukov}(2016)}]{maslov2017}%
  \BibitemOpen
  \bibfield  {author} {\bibinfo {author} {\bibfnamefont {D.~L.}\ \bibnamefont
  {Maslov}}\ and\ \bibinfo {author} {\bibfnamefont {A.~V.}\ \bibnamefont
  {Chubukov}},\ }\bibfield  {title} {\bibinfo {title} {{Optical response of
  correlated electron systems}},\ }\href
  {https://doi.org/10.1088/1361-6633/80/2/026503} {\bibfield  {journal}
  {\bibinfo  {journal} {Reports on Progress in Physics}\ }\textbf {\bibinfo
  {volume} {80}},\ \bibinfo {pages} {026503} (\bibinfo {year}
  {2016})}\BibitemShut {NoStop}%
\bibitem [{\citenamefont {Vu\ifmmode \check{c}\else \v{c}\fi{}i\ifmmode
  \check{c}\else \v{c}\fi{}evi\ifmmode~\acute{c}\else \'{c}\fi{}}\ \emph
  {et~al.}(2019)\citenamefont {Vu\ifmmode \check{c}\else \v{c}\fi{}i\ifmmode
  \check{c}\else \v{c}\fi{}evi\ifmmode~\acute{c}\else \'{c}\fi{}},
  \citenamefont {Kokalj}, \citenamefont {\ifmmode~\check{Z}\else
  \v{Z}\fi{}itko}, \citenamefont {Wentzell}, \citenamefont
  {Tanaskovi\ifmmode~\acute{c}\else \'{c}\fi{}},\ and\ \citenamefont
  {Mravlje}}]{vucicevic2019}%
  \BibitemOpen
  \bibfield  {author} {\bibinfo {author} {\bibfnamefont {J.}~\bibnamefont
  {Vu\ifmmode \check{c}\else \v{c}\fi{}i\ifmmode \check{c}\else
  \v{c}\fi{}evi\ifmmode~\acute{c}\else \'{c}\fi{}}}, \bibinfo {author}
  {\bibfnamefont {J.}~\bibnamefont {Kokalj}}, \bibinfo {author} {\bibfnamefont
  {R.}~\bibnamefont {\ifmmode~\check{Z}\else \v{Z}\fi{}itko}}, \bibinfo
  {author} {\bibfnamefont {N.}~\bibnamefont {Wentzell}}, \bibinfo {author}
  {\bibfnamefont {D.}~\bibnamefont {Tanaskovi\ifmmode~\acute{c}\else
  \'{c}\fi{}}},\ and\ \bibinfo {author} {\bibfnamefont {J.}~\bibnamefont
  {Mravlje}},\ }\bibfield  {title} {\bibinfo {title} {{Conductivity in the
  Square Lattice Hubbard Model at High Temperatures: Importance of Vertex
  Corrections}},\ }\href {https://doi.org/10.1103/PhysRevLett.123.036601}
  {\bibfield  {journal} {\bibinfo  {journal} {Phys. Rev. Lett.}\ }\textbf
  {\bibinfo {volume} {123}},\ \bibinfo {pages} {036601} (\bibinfo {year}
  {2019})}\BibitemShut {NoStop}%
\bibitem [{\citenamefont {Huang}\ \emph {et~al.}(2019)\citenamefont {Huang},
  \citenamefont {Sheppard}, \citenamefont {Moritz},\ and\ \citenamefont
  {Devereaux}}]{huang2019}%
  \BibitemOpen
  \bibfield  {author} {\bibinfo {author} {\bibfnamefont {E.~W.}\ \bibnamefont
  {Huang}}, \bibinfo {author} {\bibfnamefont {R.}~\bibnamefont {Sheppard}},
  \bibinfo {author} {\bibfnamefont {B.}~\bibnamefont {Moritz}},\ and\ \bibinfo
  {author} {\bibfnamefont {T.~P.}\ \bibnamefont {Devereaux}},\ }\bibfield
  {title} {\bibinfo {title} {{Strange metallicity in the doped Hubbard
  model}},\ }\href {https://doi.org/10.1126/science.aau7063} {\bibfield
  {journal} {\bibinfo  {journal} {Science}\ }\textbf {\bibinfo {volume}
  {366}},\ \bibinfo {pages} {987} (\bibinfo {year} {2019})}\BibitemShut
  {NoStop}%
\bibitem [{\citenamefont {Pudleiner}\ \emph {et~al.}(2019)\citenamefont
  {Pudleiner}, \citenamefont {Thunstr\"om}, \citenamefont {Valli},
  \citenamefont {Kauch}, \citenamefont {Li},\ and\ \citenamefont
  {Held}}]{pudleiner2019}%
  \BibitemOpen
  \bibfield  {author} {\bibinfo {author} {\bibfnamefont {P.}~\bibnamefont
  {Pudleiner}}, \bibinfo {author} {\bibfnamefont {P.}~\bibnamefont
  {Thunstr\"om}}, \bibinfo {author} {\bibfnamefont {A.}~\bibnamefont {Valli}},
  \bibinfo {author} {\bibfnamefont {A.}~\bibnamefont {Kauch}}, \bibinfo
  {author} {\bibfnamefont {G.}~\bibnamefont {Li}},\ and\ \bibinfo {author}
  {\bibfnamefont {K.}~\bibnamefont {Held}},\ }\bibfield  {title} {\bibinfo
  {title} {{Parquet approximation for molecules: Spectrum and optical
  conductivity of the Pariser-Parr-Pople model}},\ }\href
  {https://doi.org/10.1103/PhysRevB.99.125111} {\bibfield  {journal} {\bibinfo
  {journal} {Phys. Rev. B}\ }\textbf {\bibinfo {volume} {99}},\ \bibinfo
  {pages} {125111} (\bibinfo {year} {2019})}\BibitemShut {NoStop}%
\bibitem [{\citenamefont {Kauch}\ \emph {et~al.}(2020)\citenamefont {Kauch},
  \citenamefont {Pudleiner}, \citenamefont {Astleithner}, \citenamefont
  {Thunstr\"om}, \citenamefont {Ribic},\ and\ \citenamefont
  {Held}}]{kauch2020}%
  \BibitemOpen
  \bibfield  {author} {\bibinfo {author} {\bibfnamefont {A.}~\bibnamefont
  {Kauch}}, \bibinfo {author} {\bibfnamefont {P.}~\bibnamefont {Pudleiner}},
  \bibinfo {author} {\bibfnamefont {K.}~\bibnamefont {Astleithner}}, \bibinfo
  {author} {\bibfnamefont {P.}~\bibnamefont {Thunstr\"om}}, \bibinfo {author}
  {\bibfnamefont {T.}~\bibnamefont {Ribic}},\ and\ \bibinfo {author}
  {\bibfnamefont {K.}~\bibnamefont {Held}},\ }\bibfield  {title} {\bibinfo
  {title} {Generic optical excitations of correlated systems:
  $\ensuremath{\pi}$-tons},\ }\href
  {https://doi.org/10.1103/PhysRevLett.124.047401} {\bibfield  {journal}
  {\bibinfo  {journal} {Phys. Rev. Lett.}\ }\textbf {\bibinfo {volume} {124}},\
  \bibinfo {pages} {047401} (\bibinfo {year} {2020})}\BibitemShut {NoStop}%
\bibitem [{\citenamefont {Astleithner}\ \emph {et~al.}(2020)\citenamefont
  {Astleithner}, \citenamefont {Kauch}, \citenamefont {Ribic},\ and\
  \citenamefont {Held}}]{astleithner2020}%
  \BibitemOpen
  \bibfield  {author} {\bibinfo {author} {\bibfnamefont {K.}~\bibnamefont
  {Astleithner}}, \bibinfo {author} {\bibfnamefont {A.}~\bibnamefont {Kauch}},
  \bibinfo {author} {\bibfnamefont {T.}~\bibnamefont {Ribic}},\ and\ \bibinfo
  {author} {\bibfnamefont {K.}~\bibnamefont {Held}},\ }\bibfield  {title}
  {\bibinfo {title} {{Parquet dual fermion approach for the Falicov-Kimball
  model}},\ }\href {https://doi.org/10.1103/PhysRevB.101.165101} {\bibfield
  {journal} {\bibinfo  {journal} {Phys. Rev. B}\ }\textbf {\bibinfo {volume}
  {101}},\ \bibinfo {pages} {165101} (\bibinfo {year} {2020})}\BibitemShut
  {NoStop}%
\bibitem [{\citenamefont {Bickers}(2004)}]{bickers2004}%
  \BibitemOpen
  \bibfield  {author} {\bibinfo {author} {\bibfnamefont {N.~E.}\ \bibnamefont
  {Bickers}},\ }\bibinfo {title} {{"Self-Consistent Many-Body Theory for
  Condensed Matter Systems"}},\ in\ \href
  {https://doi.org/10.1007/0-387-21717-7_6} {\emph {\bibinfo {booktitle}
  {Theoretical Methods for Strongly Correlated Electrons}}},\ \bibinfo {editor}
  {edited by\ \bibinfo {editor} {\bibfnamefont {D.}~\bibnamefont
  {S{\'e}n{\'e}chal}}, \bibinfo {editor} {\bibfnamefont {A.-M.}\ \bibnamefont
  {Tremblay}},\ and\ \bibinfo {editor} {\bibfnamefont {C.}~\bibnamefont
  {Bourbonnais}}}\ (\bibinfo  {publisher} {Springer New York},\ \bibinfo
  {address} {New York, NY},\ \bibinfo {year} {2004})\ pp.\ \bibinfo {pages}
  {237--296}\BibitemShut {NoStop}%
\bibitem [{\citenamefont {Li}\ \emph {et~al.}(2016)\citenamefont {Li},
  \citenamefont {Wentzell}, \citenamefont {Pudleiner}, \citenamefont
  {Thunstr\"om},\ and\ \citenamefont {Held}}]{gang2016}%
  \BibitemOpen
  \bibfield  {author} {\bibinfo {author} {\bibfnamefont {G.}~\bibnamefont
  {Li}}, \bibinfo {author} {\bibfnamefont {N.}~\bibnamefont {Wentzell}},
  \bibinfo {author} {\bibfnamefont {P.}~\bibnamefont {Pudleiner}}, \bibinfo
  {author} {\bibfnamefont {P.}~\bibnamefont {Thunstr\"om}},\ and\ \bibinfo
  {author} {\bibfnamefont {K.}~\bibnamefont {Held}},\ }\bibfield  {title}
  {\bibinfo {title} {{Efficient implementation of the parquet equations: Role
  of the reducible vertex function and its kernel approximation}},\ }\href
  {https://doi.org/10.1103/PhysRevB.93.165103} {\bibfield  {journal} {\bibinfo
  {journal} {Phys. Rev. B}\ }\textbf {\bibinfo {volume} {93}},\ \bibinfo
  {pages} {165103} (\bibinfo {year} {2016})}\BibitemShut {NoStop}%
\bibitem [{\citenamefont {Li}\ \emph {et~al.}(2019)\citenamefont {Li},
  \citenamefont {Kauch}, \citenamefont {Pudleiner},\ and\ \citenamefont
  {Held}}]{gangli2016}%
  \BibitemOpen
  \bibfield  {author} {\bibinfo {author} {\bibfnamefont {G.}~\bibnamefont
  {Li}}, \bibinfo {author} {\bibfnamefont {A.}~\bibnamefont {Kauch}}, \bibinfo
  {author} {\bibfnamefont {P.}~\bibnamefont {Pudleiner}},\ and\ \bibinfo
  {author} {\bibfnamefont {K.}~\bibnamefont {Held}},\ }\bibfield  {title}
  {\bibinfo {title} {{The victory project v1.0: An efficient parquet equations
  solver}},\ }\href {https://doi.org/https://doi.org/10.1016/j.cpc.2019.03.008}
  {\bibfield  {journal} {\bibinfo  {journal} {Computer Physics Communications}\
  }\textbf {\bibinfo {volume} {241}},\ \bibinfo {pages} {146} (\bibinfo {year}
  {2019})}\BibitemShut {NoStop}%
\bibitem [{\citenamefont {Kusunose}(2006)}]{kusunose2006}%
  \BibitemOpen
  \bibfield  {author} {\bibinfo {author} {\bibfnamefont {H.}~\bibnamefont
  {Kusunose}},\ }\bibfield  {title} {\bibinfo {title} {{Influence of Spatial
  Correlations in Strongly Correlated Electron Systems: Extension to Dynamical
  Mean Field Approximation}},\ }\href {https://doi.org/10.1143/JPSJ.75.054713}
  {\bibfield  {journal} {\bibinfo  {journal} {Journal of the Physical Society
  of Japan}\ }\textbf {\bibinfo {volume} {75}},\ \bibinfo {pages} {054713}
  (\bibinfo {year} {2006})}\BibitemShut {NoStop}%
\bibitem [{\citenamefont {Toschi}\ \emph {et~al.}(2007)\citenamefont {Toschi},
  \citenamefont {Katanin},\ and\ \citenamefont {Held}}]{toschi2007}%
  \BibitemOpen
  \bibfield  {author} {\bibinfo {author} {\bibfnamefont {A.}~\bibnamefont
  {Toschi}}, \bibinfo {author} {\bibfnamefont {A.~A.}\ \bibnamefont
  {Katanin}},\ and\ \bibinfo {author} {\bibfnamefont {K.}~\bibnamefont
  {Held}},\ }\bibfield  {title} {\bibinfo {title} {{Dynamical vertex
  approximation: A step beyond dynamical mean-field theory}},\ }\href
  {https://doi.org/10.1103/PhysRevB.75.045118} {\bibfield  {journal} {\bibinfo
  {journal} {Phys. Rev. B}\ }\textbf {\bibinfo {volume} {75}},\ \bibinfo
  {pages} {045118} (\bibinfo {year} {2007})}\BibitemShut {NoStop}%
\bibitem [{\citenamefont {Katanin}\ \emph {et~al.}(2009)\citenamefont
  {Katanin}, \citenamefont {Toschi},\ and\ \citenamefont {Held}}]{katanin2009}%
  \BibitemOpen
  \bibfield  {author} {\bibinfo {author} {\bibfnamefont {A.~A.}\ \bibnamefont
  {Katanin}}, \bibinfo {author} {\bibfnamefont {A.}~\bibnamefont {Toschi}},\
  and\ \bibinfo {author} {\bibfnamefont {K.}~\bibnamefont {Held}},\ }\bibfield
  {title} {\bibinfo {title} {{Comparing pertinent effects of antiferromagnetic
  fluctuations in the two- and three-dimensional Hubbard model}},\ }\href
  {https://doi.org/10.1103/PhysRevB.80.075104} {\bibfield  {journal} {\bibinfo
  {journal} {Phys. Rev. B}\ }\textbf {\bibinfo {volume} {80}},\ \bibinfo
  {pages} {075104} (\bibinfo {year} {2009})}\BibitemShut {NoStop}%
\bibitem [{\citenamefont {Falicov}\ and\ \citenamefont
  {Kimball}(1969)}]{falicov1969}%
  \BibitemOpen
  \bibfield  {author} {\bibinfo {author} {\bibfnamefont {L.~M.}\ \bibnamefont
  {Falicov}}\ and\ \bibinfo {author} {\bibfnamefont {J.~C.}\ \bibnamefont
  {Kimball}},\ }\bibfield  {title} {\bibinfo {title} {{Simple Model for
  Semiconductor-Metal Transitions: Sm${\mathrm{B}}_{6}$ and Transition-Metal
  Oxides}},\ }\href {https://doi.org/10.1103/PhysRevLett.22.997} {\bibfield
  {journal} {\bibinfo  {journal} {Phys. Rev. Lett.}\ }\textbf {\bibinfo
  {volume} {22}},\ \bibinfo {pages} {997} (\bibinfo {year} {1969})}\BibitemShut
  {NoStop}%
\bibitem [{\citenamefont {Simard}\ \emph
  {et~al.}(2021{\natexlab{a}})\citenamefont {Simard}, \citenamefont
  {Takayoshi},\ and\ \citenamefont {Werner}}]{simard_eq}%
  \BibitemOpen
  \bibfield  {author} {\bibinfo {author} {\bibfnamefont {O.}~\bibnamefont
  {Simard}}, \bibinfo {author} {\bibfnamefont {S.}~\bibnamefont {Takayoshi}},\
  and\ \bibinfo {author} {\bibfnamefont {P.}~\bibnamefont {Werner}},\
  }\bibfield  {title} {\bibinfo {title} {{Diagrammatic study of optical
  excitations in correlated systems}},\ }\href
  {https://doi.org/10.1103/PhysRevB.103.104415} {\bibfield  {journal} {\bibinfo
   {journal} {Phys. Rev. B}\ }\textbf {\bibinfo {volume} {103}},\ \bibinfo
  {pages} {104415} (\bibinfo {year} {2021}{\natexlab{a}})}\BibitemShut
  {NoStop}%
\bibitem [{\citenamefont {Simard}\ \emph
  {et~al.}(2021{\natexlab{b}})\citenamefont {Simard}, \citenamefont
  {Eckstein},\ and\ \citenamefont {Werner}}]{simard_noneq}%
  \BibitemOpen
  \bibfield  {author} {\bibinfo {author} {\bibfnamefont {O.}~\bibnamefont
  {Simard}}, \bibinfo {author} {\bibfnamefont {M.}~\bibnamefont {Eckstein}},\
  and\ \bibinfo {author} {\bibfnamefont {P.}~\bibnamefont {Werner}},\
  }\bibfield  {title} {\bibinfo {title} {{Nonequilibrium evolution of the
  optical conductivity of the weakly interacting Hubbard model: Drude response
  and $\ensuremath{\pi}$-ton type vertex corrections}},\ }\href
  {https://doi.org/10.1103/PhysRevB.104.245127} {\bibfield  {journal} {\bibinfo
   {journal} {Phys. Rev. B}\ }\textbf {\bibinfo {volume} {104}},\ \bibinfo
  {pages} {245127} (\bibinfo {year} {2021}{\natexlab{b}})}\BibitemShut
  {NoStop}%
\bibitem [{\citenamefont {Worm}\ \emph {et~al.}(2021)\citenamefont {Worm},
  \citenamefont {Watzenb\"ock}, \citenamefont {Pickem}, \citenamefont {Kauch},\
  and\ \citenamefont {Held}}]{worm2021}%
  \BibitemOpen
  \bibfield  {author} {\bibinfo {author} {\bibfnamefont {P.}~\bibnamefont
  {Worm}}, \bibinfo {author} {\bibfnamefont {C.}~\bibnamefont {Watzenb\"ock}},
  \bibinfo {author} {\bibfnamefont {M.}~\bibnamefont {Pickem}}, \bibinfo
  {author} {\bibfnamefont {A.}~\bibnamefont {Kauch}},\ and\ \bibinfo {author}
  {\bibfnamefont {K.}~\bibnamefont {Held}},\ }\bibfield  {title} {\bibinfo
  {title} {{Broadening and sharpening of the Drude peak through
  antiferromagnetic fluctuations}},\ }\href
  {https://doi.org/10.1103/PhysRevB.104.115153} {\bibfield  {journal} {\bibinfo
   {journal} {Phys. Rev. B}\ }\textbf {\bibinfo {volume} {104}},\ \bibinfo
  {pages} {115153} (\bibinfo {year} {2021})}\BibitemShut {NoStop}%
\bibitem [{\citenamefont {Rozenberg}\ \emph {et~al.}(1995)\citenamefont
  {Rozenberg}, \citenamefont {Kotliar}, \citenamefont {Kajueter}, \citenamefont
  {Thomas}, \citenamefont {Rapkine}, \citenamefont {Honig},\ and\ \citenamefont
  {Metcalf}}]{rozenberg1995}%
  \BibitemOpen
  \bibfield  {author} {\bibinfo {author} {\bibfnamefont {M.~J.}\ \bibnamefont
  {Rozenberg}}, \bibinfo {author} {\bibfnamefont {G.}~\bibnamefont {Kotliar}},
  \bibinfo {author} {\bibfnamefont {H.}~\bibnamefont {Kajueter}}, \bibinfo
  {author} {\bibfnamefont {G.~A.}\ \bibnamefont {Thomas}}, \bibinfo {author}
  {\bibfnamefont {D.~H.}\ \bibnamefont {Rapkine}}, \bibinfo {author}
  {\bibfnamefont {J.~M.}\ \bibnamefont {Honig}},\ and\ \bibinfo {author}
  {\bibfnamefont {P.}~\bibnamefont {Metcalf}},\ }\bibfield  {title} {\bibinfo
  {title} {{Optical Conductivity in Mott-Hubbard Systems}},\ }\href
  {https://doi.org/10.1103/PhysRevLett.75.105} {\bibfield  {journal} {\bibinfo
  {journal} {Phys. Rev. Lett.}\ }\textbf {\bibinfo {volume} {75}},\ \bibinfo
  {pages} {105} (\bibinfo {year} {1995})}\BibitemShut {NoStop}%
\bibitem [{\citenamefont {Puchkov}\ \emph {et~al.}(1995)\citenamefont
  {Puchkov}, \citenamefont {Timusk}, \citenamefont {Doyle},\ and\ \citenamefont
  {Hermann}}]{puchkov1995}%
  \BibitemOpen
  \bibfield  {author} {\bibinfo {author} {\bibfnamefont {A.}~\bibnamefont
  {Puchkov}}, \bibinfo {author} {\bibfnamefont {T.}~\bibnamefont {Timusk}},
  \bibinfo {author} {\bibfnamefont {S.}~\bibnamefont {Doyle}},\ and\ \bibinfo
  {author} {\bibfnamefont {A.}~\bibnamefont {Hermann}},\ }\bibfield  {title}
  {\bibinfo {title} {{ab-plane optical properties of
  ${\mathrm{Tl}}_{2}$${\mathrm{Ba}}_{2}$${\mathrm{CuO}}_{6+\mathrm{\ensuremath{\delta}}}$}},\
  }\href {https://doi.org/10.1103/PhysRevB.51.3312} {\bibfield  {journal}
  {\bibinfo  {journal} {Phys. Rev. B}\ }\textbf {\bibinfo {volume} {51}},\
  \bibinfo {pages} {3312} (\bibinfo {year} {1995})}\BibitemShut {NoStop}%
\bibitem [{\citenamefont {Tsvetkov}\ \emph {et~al.}(1997)\citenamefont
  {Tsvetkov}, \citenamefont {Sch\"utzmann}, \citenamefont {Gorina},
  \citenamefont {Kaljushnaia},\ and\ \citenamefont {van~der
  Marel}}]{tsvetkov1997}%
  \BibitemOpen
  \bibfield  {author} {\bibinfo {author} {\bibfnamefont {A.~A.}\ \bibnamefont
  {Tsvetkov}}, \bibinfo {author} {\bibfnamefont {J.}~\bibnamefont
  {Sch\"utzmann}}, \bibinfo {author} {\bibfnamefont {J.~I.}\ \bibnamefont
  {Gorina}}, \bibinfo {author} {\bibfnamefont {G.~A.}\ \bibnamefont
  {Kaljushnaia}},\ and\ \bibinfo {author} {\bibfnamefont {D.}~\bibnamefont
  {van~der Marel}},\ }\bibfield  {title} {\bibinfo {title} {{In-plane optical
  response of ${\mathrm{Bi}}_{2}$${\mathrm{Sr}}_{2}$${\mathrm{CuO}}_{6}$}},\
  }\href {https://doi.org/10.1103/PhysRevB.55.14152} {\bibfield  {journal}
  {\bibinfo  {journal} {Phys. Rev. B}\ }\textbf {\bibinfo {volume} {55}},\
  \bibinfo {pages} {14152} (\bibinfo {year} {1997})}\BibitemShut {NoStop}%
\bibitem [{\citenamefont {Wang}\ \emph {et~al.}(1998)\citenamefont {Wang},
  \citenamefont {Tajima}, \citenamefont {Rykov},\ and\ \citenamefont
  {Tomimoto}}]{wang1998}%
  \BibitemOpen
  \bibfield  {author} {\bibinfo {author} {\bibfnamefont {N.~L.}\ \bibnamefont
  {Wang}}, \bibinfo {author} {\bibfnamefont {S.}~\bibnamefont {Tajima}},
  \bibinfo {author} {\bibfnamefont {A.~I.}\ \bibnamefont {Rykov}},\ and\
  \bibinfo {author} {\bibfnamefont {K.}~\bibnamefont {Tomimoto}},\ }\bibfield
  {title} {\bibinfo {title} {{Zn-substitution effects on the optical
  conductivity in
  ${\mathrm{YBa}}_{2}{\mathrm{Cu}}_{3}{\mathrm{O}}_{7\mathrm{\ensuremath{-}}\mathrm{\ensuremath{\delta}}}$
  crystals: Strong pair breaking and reduction of in-plane anisotropy}},\
  }\href {https://doi.org/10.1103/PhysRevB.57.R11081} {\bibfield  {journal}
  {\bibinfo  {journal} {Phys. Rev. B}\ }\textbf {\bibinfo {volume} {57}},\
  \bibinfo {pages} {R11081} (\bibinfo {year} {1998})}\BibitemShut {NoStop}%
\bibitem [{\citenamefont {Osafune}\ \emph {et~al.}(1999)\citenamefont
  {Osafune}, \citenamefont {Motoyama}, \citenamefont {Eisaki}, \citenamefont
  {Uchida},\ and\ \citenamefont {Tajima}}]{osafune1999}%
  \BibitemOpen
  \bibfield  {author} {\bibinfo {author} {\bibfnamefont {T.}~\bibnamefont
  {Osafune}}, \bibinfo {author} {\bibfnamefont {N.}~\bibnamefont {Motoyama}},
  \bibinfo {author} {\bibfnamefont {H.}~\bibnamefont {Eisaki}}, \bibinfo
  {author} {\bibfnamefont {S.}~\bibnamefont {Uchida}},\ and\ \bibinfo {author}
  {\bibfnamefont {S.}~\bibnamefont {Tajima}},\ }\bibfield  {title} {\bibinfo
  {title} {{Pseudogap and Collective Mode in the Optical Conductivity Spectra
  of Hole-Doped Ladders in
  ${\mathrm{Sr}}_{14\ensuremath{-}\mathit{x}}{\mathrm{Ca}}_{\mathit{x}}{\mathrm{Cu}}_{24}{O}_{41}$}},\
  }\href {https://doi.org/10.1103/PhysRevLett.82.1313} {\bibfield  {journal}
  {\bibinfo  {journal} {Phys. Rev. Lett.}\ }\textbf {\bibinfo {volume} {82}},\
  \bibinfo {pages} {1313} (\bibinfo {year} {1999})}\BibitemShut {NoStop}%
\bibitem [{\citenamefont {Takenaka}\ \emph {et~al.}(1999)\citenamefont
  {Takenaka}, \citenamefont {Sawaki},\ and\ \citenamefont
  {Sugai}}]{takenaka1999}%
  \BibitemOpen
  \bibfield  {author} {\bibinfo {author} {\bibfnamefont {K.}~\bibnamefont
  {Takenaka}}, \bibinfo {author} {\bibfnamefont {Y.}~\bibnamefont {Sawaki}},\
  and\ \bibinfo {author} {\bibfnamefont {S.}~\bibnamefont {Sugai}},\ }\bibfield
   {title} {\bibinfo {title} {{Incoherent-to-coherent crossover of optical
  spectra in ${\mathrm{La}}_{0.825}{\mathrm{Sr}}_{0.175}{\mathrm{MnO}}_{3}:$
  Temperature-dependent reflectivity spectra measured on cleaved surfaces}},\
  }\href {https://doi.org/10.1103/PhysRevB.60.13011} {\bibfield  {journal}
  {\bibinfo  {journal} {Phys. Rev. B}\ }\textbf {\bibinfo {volume} {60}},\
  \bibinfo {pages} {13011} (\bibinfo {year} {1999})}\BibitemShut {NoStop}%
\bibitem [{\citenamefont {Lupi}\ \emph {et~al.}(2000)\citenamefont {Lupi},
  \citenamefont {Calvani}, \citenamefont {Capizzi},\ and\ \citenamefont
  {Roy}}]{lupi2000}%
  \BibitemOpen
  \bibfield  {author} {\bibinfo {author} {\bibfnamefont {S.}~\bibnamefont
  {Lupi}}, \bibinfo {author} {\bibfnamefont {P.}~\bibnamefont {Calvani}},
  \bibinfo {author} {\bibfnamefont {M.}~\bibnamefont {Capizzi}},\ and\ \bibinfo
  {author} {\bibfnamefont {P.}~\bibnamefont {Roy}},\ }\bibfield  {title}
  {\bibinfo {title} {{Evidence of two species of carriers from the far-infrared
  reflectivity of ${\mathbf{Bi}}_{2}{\mathbf{Sr}}_{2}{\mathbf{CuO}}_{6}$}},\
  }\href {https://doi.org/10.1103/PhysRevB.62.12418} {\bibfield  {journal}
  {\bibinfo  {journal} {Phys. Rev. B}\ }\textbf {\bibinfo {volume} {62}},\
  \bibinfo {pages} {12418} (\bibinfo {year} {2000})}\BibitemShut {NoStop}%
\bibitem [{\citenamefont {Kostic}\ \emph {et~al.}(1998)\citenamefont {Kostic},
  \citenamefont {Okada}, \citenamefont {Collins}, \citenamefont {Schlesinger},
  \citenamefont {Reiner}, \citenamefont {Klein}, \citenamefont {Kapitulnik},
  \citenamefont {Geballe},\ and\ \citenamefont {Beasley}}]{kostic1998}%
  \BibitemOpen
  \bibfield  {author} {\bibinfo {author} {\bibfnamefont {P.}~\bibnamefont
  {Kostic}}, \bibinfo {author} {\bibfnamefont {Y.}~\bibnamefont {Okada}},
  \bibinfo {author} {\bibfnamefont {N.~C.}\ \bibnamefont {Collins}}, \bibinfo
  {author} {\bibfnamefont {Z.}~\bibnamefont {Schlesinger}}, \bibinfo {author}
  {\bibfnamefont {J.~W.}\ \bibnamefont {Reiner}}, \bibinfo {author}
  {\bibfnamefont {L.}~\bibnamefont {Klein}}, \bibinfo {author} {\bibfnamefont
  {A.}~\bibnamefont {Kapitulnik}}, \bibinfo {author} {\bibfnamefont {T.~H.}\
  \bibnamefont {Geballe}},\ and\ \bibinfo {author} {\bibfnamefont {M.~R.}\
  \bibnamefont {Beasley}},\ }\bibfield  {title} {\bibinfo {title}
  {{Non-Fermi-Liquid Behavior of $\mathrm{SrRuO}{}_{3}$: Evidence from Infrared
  Conductivity}},\ }\href {https://doi.org/10.1103/PhysRevLett.81.2498}
  {\bibfield  {journal} {\bibinfo  {journal} {Phys. Rev. Lett.}\ }\textbf
  {\bibinfo {volume} {81}},\ \bibinfo {pages} {2498} (\bibinfo {year}
  {1998})}\BibitemShut {NoStop}%
\bibitem [{\citenamefont {Lee}\ \emph {et~al.}(2002)\citenamefont {Lee},
  \citenamefont {Yu}, \citenamefont {Lee}, \citenamefont {Noh}, \citenamefont
  {Gimm}, \citenamefont {Choi},\ and\ \citenamefont {Eom}}]{lee2002}%
  \BibitemOpen
  \bibfield  {author} {\bibinfo {author} {\bibfnamefont {Y.~S.}\ \bibnamefont
  {Lee}}, \bibinfo {author} {\bibfnamefont {J.}~\bibnamefont {Yu}}, \bibinfo
  {author} {\bibfnamefont {J.~S.}\ \bibnamefont {Lee}}, \bibinfo {author}
  {\bibfnamefont {T.~W.}\ \bibnamefont {Noh}}, \bibinfo {author} {\bibfnamefont
  {T.-H.}\ \bibnamefont {Gimm}}, \bibinfo {author} {\bibfnamefont {H.-Y.}\
  \bibnamefont {Choi}},\ and\ \bibinfo {author} {\bibfnamefont {C.~B.}\
  \bibnamefont {Eom}},\ }\bibfield  {title} {\bibinfo {title} {{Non-Fermi
  liquid behavior and scaling of the low-frequency suppression in the optical
  conductivity spectra of ${\mathrm{CaRuO}}_{3}$}},\ }\href
  {https://doi.org/10.1103/PhysRevB.66.041104} {\bibfield  {journal} {\bibinfo
  {journal} {Phys. Rev. B}\ }\textbf {\bibinfo {volume} {66}},\ \bibinfo
  {pages} {041104} (\bibinfo {year} {2002})}\BibitemShut {NoStop}%
\bibitem [{\citenamefont {Takenaka}\ \emph {et~al.}(2002)\citenamefont
  {Takenaka}, \citenamefont {Shiozaki},\ and\ \citenamefont
  {Sugai}}]{takenaka2002}%
  \BibitemOpen
  \bibfield  {author} {\bibinfo {author} {\bibfnamefont {K.}~\bibnamefont
  {Takenaka}}, \bibinfo {author} {\bibfnamefont {R.}~\bibnamefont {Shiozaki}},\
  and\ \bibinfo {author} {\bibfnamefont {S.}~\bibnamefont {Sugai}},\ }\bibfield
   {title} {\bibinfo {title} {{Charge dynamics of a double-exchange ferromagnet
  ${\mathrm{La}}_{1\ensuremath{-}x}{\mathrm{Sr}}_{x}{\mathrm{MnO}}_{3}$}},\
  }\href {https://doi.org/10.1103/PhysRevB.65.184436} {\bibfield  {journal}
  {\bibinfo  {journal} {Phys. Rev. B}\ }\textbf {\bibinfo {volume} {65}},\
  \bibinfo {pages} {184436} (\bibinfo {year} {2002})}\BibitemShut {NoStop}%
\bibitem [{\citenamefont {Santander-Syro}\ \emph {et~al.}(2002)\citenamefont
  {Santander-Syro}, \citenamefont {Lobo}, \citenamefont {Bontemps},
  \citenamefont {Konstantinovic}, \citenamefont {Li},\ and\ \citenamefont
  {Raffy}}]{santandersyro2002}%
  \BibitemOpen
  \bibfield  {author} {\bibinfo {author} {\bibfnamefont {A.~F.}\ \bibnamefont
  {Santander-Syro}}, \bibinfo {author} {\bibfnamefont {R.~P. S.~M.}\
  \bibnamefont {Lobo}}, \bibinfo {author} {\bibfnamefont {N.}~\bibnamefont
  {Bontemps}}, \bibinfo {author} {\bibfnamefont {Z.}~\bibnamefont
  {Konstantinovic}}, \bibinfo {author} {\bibfnamefont {Z.}~\bibnamefont {Li}},\
  and\ \bibinfo {author} {\bibfnamefont {H.}~\bibnamefont {Raffy}},\ }\bibfield
   {title} {\bibinfo {title} {{Absence of a Loss of In-Plane Infrared Spectral
  Weight in the Pseudogap Regime of
  ${\mathrm{Bi}}_{2}{\mathrm{Sr}}_{2}{\mathrm{CaCu}}_{2}{O}_{8+\ensuremath{\delta}}$}},\
  }\href {https://doi.org/10.1103/PhysRevLett.88.097005} {\bibfield  {journal}
  {\bibinfo  {journal} {Phys. Rev. Lett.}\ }\textbf {\bibinfo {volume} {88}},\
  \bibinfo {pages} {097005} (\bibinfo {year} {2002})}\BibitemShut {NoStop}%
\bibitem [{\citenamefont {Takenaka}\ \emph {et~al.}(2003)\citenamefont
  {Takenaka}, \citenamefont {Nohara}, \citenamefont {Shiozaki},\ and\
  \citenamefont {Sugai}}]{takenaka2003}%
  \BibitemOpen
  \bibfield  {author} {\bibinfo {author} {\bibfnamefont {K.}~\bibnamefont
  {Takenaka}}, \bibinfo {author} {\bibfnamefont {J.}~\bibnamefont {Nohara}},
  \bibinfo {author} {\bibfnamefont {R.}~\bibnamefont {Shiozaki}},\ and\
  \bibinfo {author} {\bibfnamefont {S.}~\bibnamefont {Sugai}},\ }\bibfield
  {title} {\bibinfo {title} {{Incoherent charge dynamics of
  ${\mathrm{La}}_{2\ensuremath{-}x}{\mathrm{Sr}}_{x}{\mathrm{CuO}}_{4}:$
  Dynamical localization and resistivity saturation}},\ }\href
  {https://doi.org/10.1103/PhysRevB.68.134501} {\bibfield  {journal} {\bibinfo
  {journal} {Phys. Rev. B}\ }\textbf {\bibinfo {volume} {68}},\ \bibinfo
  {pages} {134501} (\bibinfo {year} {2003})}\BibitemShut {NoStop}%
\bibitem [{\citenamefont {Wang}\ \emph {et~al.}(2003)\citenamefont {Wang},
  \citenamefont {Zheng}, \citenamefont {Feng}, \citenamefont {Gu},
  \citenamefont {Homes}, \citenamefont {Tranquada}, \citenamefont {Gaulin},\
  and\ \citenamefont {Timusk}}]{wang2003}%
  \BibitemOpen
  \bibfield  {author} {\bibinfo {author} {\bibfnamefont {N.~L.}\ \bibnamefont
  {Wang}}, \bibinfo {author} {\bibfnamefont {P.}~\bibnamefont {Zheng}},
  \bibinfo {author} {\bibfnamefont {T.}~\bibnamefont {Feng}}, \bibinfo {author}
  {\bibfnamefont {G.~D.}\ \bibnamefont {Gu}}, \bibinfo {author} {\bibfnamefont
  {C.~C.}\ \bibnamefont {Homes}}, \bibinfo {author} {\bibfnamefont {J.~M.}\
  \bibnamefont {Tranquada}}, \bibinfo {author} {\bibfnamefont {B.~D.}\
  \bibnamefont {Gaulin}},\ and\ \bibinfo {author} {\bibfnamefont
  {T.}~\bibnamefont {Timusk}},\ }\bibfield  {title} {\bibinfo {title}
  {{Infrared properties of
  ${\mathrm{La}}_{2\ensuremath{-}x}(\mathrm{C}\mathrm{a},\mathrm{S}\mathrm{r}{)}_{x}{\mathrm{CaCu}}_{2}{\mathrm{O}}_{6+\ensuremath{\delta}}$
  single crystals}},\ }\href {https://doi.org/10.1103/PhysRevB.67.134526}
  {\bibfield  {journal} {\bibinfo  {journal} {Phys. Rev. B}\ }\textbf {\bibinfo
  {volume} {67}},\ \bibinfo {pages} {134526} (\bibinfo {year}
  {2003})}\BibitemShut {NoStop}%
\bibitem [{\citenamefont {N.~E. Hussey~‖}\ and\ \citenamefont
  {Takagi}(2004)}]{hussey2004}%
  \BibitemOpen
  \bibfield  {author} {\bibinfo {author} {\bibfnamefont {K.~T.}\ \bibnamefont
  {N.~E. Hussey~‖}}\ and\ \bibinfo {author} {\bibfnamefont {H.}~\bibnamefont
  {Takagi}},\ }\bibfield  {title} {\bibinfo {title} {{Universality of the
  Mott–Ioffe–Regel limit in metals}},\ }\href
  {https://doi.org/10.1080/14786430410001716944} {\bibfield  {journal}
  {\bibinfo  {journal} {Philosophical Magazine}\ }\textbf {\bibinfo {volume}
  {84}},\ \bibinfo {pages} {2847} (\bibinfo {year} {2004})}\BibitemShut
  {NoStop}%
\bibitem [{\citenamefont {Wang}\ \emph {et~al.}(2004)\citenamefont {Wang},
  \citenamefont {Zheng}, \citenamefont {Wu}, \citenamefont {Ma}, \citenamefont
  {Xiang}, \citenamefont {Jin},\ and\ \citenamefont {Mandrus}}]{wang2004}%
  \BibitemOpen
  \bibfield  {author} {\bibinfo {author} {\bibfnamefont {N.~L.}\ \bibnamefont
  {Wang}}, \bibinfo {author} {\bibfnamefont {P.}~\bibnamefont {Zheng}},
  \bibinfo {author} {\bibfnamefont {D.}~\bibnamefont {Wu}}, \bibinfo {author}
  {\bibfnamefont {Y.~C.}\ \bibnamefont {Ma}}, \bibinfo {author} {\bibfnamefont
  {T.}~\bibnamefont {Xiang}}, \bibinfo {author} {\bibfnamefont {R.~Y.}\
  \bibnamefont {Jin}},\ and\ \bibinfo {author} {\bibfnamefont {D.}~\bibnamefont
  {Mandrus}},\ }\bibfield  {title} {\bibinfo {title} {{Infrared Probe of the
  Electronic Structure and Charge Dynamics of
  ${\mathrm{N}\mathrm{a}}_{0.7}\mathrm{C}\mathrm{o}{\mathrm{O}}_{2}$}},\ }\href
  {https://doi.org/10.1103/PhysRevLett.93.237007} {\bibfield  {journal}
  {\bibinfo  {journal} {Phys. Rev. Lett.}\ }\textbf {\bibinfo {volume} {93}},\
  \bibinfo {pages} {237007} (\bibinfo {year} {2004})}\BibitemShut {NoStop}%
\bibitem [{\citenamefont {Takenaka}\ \emph {et~al.}(2005)\citenamefont
  {Takenaka}, \citenamefont {Tamura}, \citenamefont {Tajima}, \citenamefont
  {Takagi}, \citenamefont {Nohara},\ and\ \citenamefont
  {Sugai}}]{takenaka2005}%
  \BibitemOpen
  \bibfield  {author} {\bibinfo {author} {\bibfnamefont {K.}~\bibnamefont
  {Takenaka}}, \bibinfo {author} {\bibfnamefont {M.}~\bibnamefont {Tamura}},
  \bibinfo {author} {\bibfnamefont {N.}~\bibnamefont {Tajima}}, \bibinfo
  {author} {\bibfnamefont {H.}~\bibnamefont {Takagi}}, \bibinfo {author}
  {\bibfnamefont {J.}~\bibnamefont {Nohara}},\ and\ \bibinfo {author}
  {\bibfnamefont {S.}~\bibnamefont {Sugai}},\ }\bibfield  {title} {\bibinfo
  {title} {{Collapse of Coherent Quasiparticle States in
  $\ensuremath{\theta}\mathrm{\text{\ensuremath{-}}}(\mathrm{BEDT}\mathrm{\text{\ensuremath{-}}}\mathrm{TTF}{)}_{2}{\mathrm{I}}_{3}$
  Observed by Optical Spectroscopy}},\ }\href
  {https://doi.org/10.1103/PhysRevLett.95.227801} {\bibfield  {journal}
  {\bibinfo  {journal} {Phys. Rev. Lett.}\ }\textbf {\bibinfo {volume} {95}},\
  \bibinfo {pages} {227801} (\bibinfo {year} {2005})}\BibitemShut {NoStop}%
\bibitem [{\citenamefont {J\"onsson}\ \emph {et~al.}(2007)\citenamefont
  {J\"onsson}, \citenamefont {Takenaka}, \citenamefont {Niitaka}, \citenamefont
  {Sasagawa}, \citenamefont {Sugai},\ and\ \citenamefont
  {Takagi}}]{jonsson2007}%
  \BibitemOpen
  \bibfield  {author} {\bibinfo {author} {\bibfnamefont {P.~E.}\ \bibnamefont
  {J\"onsson}}, \bibinfo {author} {\bibfnamefont {K.}~\bibnamefont {Takenaka}},
  \bibinfo {author} {\bibfnamefont {S.}~\bibnamefont {Niitaka}}, \bibinfo
  {author} {\bibfnamefont {T.}~\bibnamefont {Sasagawa}}, \bibinfo {author}
  {\bibfnamefont {S.}~\bibnamefont {Sugai}},\ and\ \bibinfo {author}
  {\bibfnamefont {H.}~\bibnamefont {Takagi}},\ }\bibfield  {title} {\bibinfo
  {title} {{Correlation-Driven Heavy-Fermion Formation in
  ${\mathrm{LiV}}_{2}{\mathrm{O}}_{4}$}},\ }\href
  {https://doi.org/10.1103/PhysRevLett.99.167402} {\bibfield  {journal}
  {\bibinfo  {journal} {Phys. Rev. Lett.}\ }\textbf {\bibinfo {volume} {99}},\
  \bibinfo {pages} {167402} (\bibinfo {year} {2007})}\BibitemShut {NoStop}%
\bibitem [{\citenamefont {Kaiser}\ \emph {et~al.}(2010)\citenamefont {Kaiser},
  \citenamefont {Dressel}, \citenamefont {Sun}, \citenamefont {Greco},
  \citenamefont {Schlueter}, \citenamefont {Gard},\ and\ \citenamefont
  {Drichko}}]{kaiser2010}%
  \BibitemOpen
  \bibfield  {author} {\bibinfo {author} {\bibfnamefont {S.}~\bibnamefont
  {Kaiser}}, \bibinfo {author} {\bibfnamefont {M.}~\bibnamefont {Dressel}},
  \bibinfo {author} {\bibfnamefont {Y.}~\bibnamefont {Sun}}, \bibinfo {author}
  {\bibfnamefont {A.}~\bibnamefont {Greco}}, \bibinfo {author} {\bibfnamefont
  {J.~A.}\ \bibnamefont {Schlueter}}, \bibinfo {author} {\bibfnamefont {G.~L.}\
  \bibnamefont {Gard}},\ and\ \bibinfo {author} {\bibfnamefont
  {N.}~\bibnamefont {Drichko}},\ }\bibfield  {title} {\bibinfo {title}
  {{Bandwidth Tuning Triggers Interplay of Charge Order and Superconductivity
  in Two-Dimensional Organic Materials}},\ }\href
  {https://doi.org/10.1103/PhysRevLett.105.206402} {\bibfield  {journal}
  {\bibinfo  {journal} {Phys. Rev. Lett.}\ }\textbf {\bibinfo {volume} {105}},\
  \bibinfo {pages} {206402} (\bibinfo {year} {2010})}\BibitemShut {NoStop}%
\bibitem [{\citenamefont {Jaramillo}\ \emph {et~al.}(2014)\citenamefont
  {Jaramillo}, \citenamefont {Ha}, \citenamefont {Silevitch},\ and\
  \citenamefont {Ramanathan}}]{jaramillo2014}%
  \BibitemOpen
  \bibfield  {author} {\bibinfo {author} {\bibfnamefont {R.}~\bibnamefont
  {Jaramillo}}, \bibinfo {author} {\bibfnamefont {S.~D.}\ \bibnamefont {Ha}},
  \bibinfo {author} {\bibfnamefont {D.~M.}\ \bibnamefont {Silevitch}},\ and\
  \bibinfo {author} {\bibfnamefont {S.}~\bibnamefont {Ramanathan}},\ }\bibfield
   {title} {\bibinfo {title} {{Origins of bad-metal conductivity and the
  insulator--metal transition in the rare-earth nickelates}},\ }\href
  {https://doi.org/10.1038/nphys2907} {\bibfield  {journal} {\bibinfo
  {journal} {Nature Physics}\ }\textbf {\bibinfo {volume} {10}},\ \bibinfo
  {pages} {304} (\bibinfo {year} {2014})}\BibitemShut {NoStop}%
\bibitem [{\citenamefont {Biswas}\ \emph {et~al.}(2020)\citenamefont {Biswas},
  \citenamefont {Iakutkina}, \citenamefont {Wang}, \citenamefont {Lei},
  \citenamefont {Dressel},\ and\ \citenamefont {Uykur}}]{biswas2020}%
  \BibitemOpen
  \bibfield  {author} {\bibinfo {author} {\bibfnamefont {A.}~\bibnamefont
  {Biswas}}, \bibinfo {author} {\bibfnamefont {O.}~\bibnamefont {Iakutkina}},
  \bibinfo {author} {\bibfnamefont {Q.}~\bibnamefont {Wang}}, \bibinfo {author}
  {\bibfnamefont {H.~C.}\ \bibnamefont {Lei}}, \bibinfo {author} {\bibfnamefont
  {M.}~\bibnamefont {Dressel}},\ and\ \bibinfo {author} {\bibfnamefont
  {E.}~\bibnamefont {Uykur}},\ }\bibfield  {title} {\bibinfo {title}
  {{Spin-Reorientation-Induced Band Gap in
  ${\mathrm{Fe}}_{3}{\mathrm{Sn}}_{2}$: Optical Signatures of Weyl Nodes}},\
  }\href {https://doi.org/10.1103/PhysRevLett.125.076403} {\bibfield  {journal}
  {\bibinfo  {journal} {Phys. Rev. Lett.}\ }\textbf {\bibinfo {volume} {125}},\
  \bibinfo {pages} {076403} (\bibinfo {year} {2020})}\BibitemShut {NoStop}%
\bibitem [{\citenamefont {Pustogow}\ \emph {et~al.}(2021)\citenamefont
  {Pustogow}, \citenamefont {Saito}, \citenamefont {L{\"o}hle}, \citenamefont
  {Sanz~Alonso}, \citenamefont {Kawamoto}, \citenamefont {Dobrosavljevi{\'c}},
  \citenamefont {Dressel},\ and\ \citenamefont {Fratini}}]{pustogow2021}%
  \BibitemOpen
  \bibfield  {author} {\bibinfo {author} {\bibfnamefont {A.}~\bibnamefont
  {Pustogow}}, \bibinfo {author} {\bibfnamefont {Y.}~\bibnamefont {Saito}},
  \bibinfo {author} {\bibfnamefont {A.}~\bibnamefont {L{\"o}hle}}, \bibinfo
  {author} {\bibfnamefont {M.}~\bibnamefont {Sanz~Alonso}}, \bibinfo {author}
  {\bibfnamefont {A.}~\bibnamefont {Kawamoto}}, \bibinfo {author}
  {\bibfnamefont {V.}~\bibnamefont {Dobrosavljevi{\'c}}}, \bibinfo {author}
  {\bibfnamefont {M.}~\bibnamefont {Dressel}},\ and\ \bibinfo {author}
  {\bibfnamefont {S.}~\bibnamefont {Fratini}},\ }\bibfield  {title} {\bibinfo
  {title} {{Rise and fall of Landau's quasiparticles while approaching the Mott
  transition}},\ }\href {https://doi.org/10.1038/s41467-021-21741-z} {\bibfield
   {journal} {\bibinfo  {journal} {Nature Communications}\ }\textbf {\bibinfo
  {volume} {12}},\ \bibinfo {pages} {1571} (\bibinfo {year}
  {2021})}\BibitemShut {NoStop}%
\bibitem [{\citenamefont {Uykur}\ \emph {et~al.}(2021)\citenamefont {Uykur},
  \citenamefont {Ortiz}, \citenamefont {Iakutkina}, \citenamefont {Wenzel},
  \citenamefont {Wilson}, \citenamefont {Dressel},\ and\ \citenamefont
  {Tsirlin}}]{uzkur2021}%
  \BibitemOpen
  \bibfield  {author} {\bibinfo {author} {\bibfnamefont {E.}~\bibnamefont
  {Uykur}}, \bibinfo {author} {\bibfnamefont {B.~R.}\ \bibnamefont {Ortiz}},
  \bibinfo {author} {\bibfnamefont {O.}~\bibnamefont {Iakutkina}}, \bibinfo
  {author} {\bibfnamefont {M.}~\bibnamefont {Wenzel}}, \bibinfo {author}
  {\bibfnamefont {S.~D.}\ \bibnamefont {Wilson}}, \bibinfo {author}
  {\bibfnamefont {M.}~\bibnamefont {Dressel}},\ and\ \bibinfo {author}
  {\bibfnamefont {A.~A.}\ \bibnamefont {Tsirlin}},\ }\bibfield  {title}
  {\bibinfo {title} {{Low-energy optical properties of the nonmagnetic kagome
  metal ${\mathrm{CsV}}_{3}{\mathrm{Sb}}_{5}$}},\ }\href
  {https://doi.org/10.1103/PhysRevB.104.045130} {\bibfield  {journal} {\bibinfo
   {journal} {Phys. Rev. B}\ }\textbf {\bibinfo {volume} {104}},\ \bibinfo
  {pages} {045130} (\bibinfo {year} {2021})}\BibitemShut {NoStop}%
\bibitem [{\citenamefont {Uykur}\ \emph {et~al.}(2022)\citenamefont {Uykur},
  \citenamefont {Ortiz}, \citenamefont {Wilson}, \citenamefont {Dressel},\ and\
  \citenamefont {Tsirlin}}]{uykur2022}%
  \BibitemOpen
  \bibfield  {author} {\bibinfo {author} {\bibfnamefont {E.}~\bibnamefont
  {Uykur}}, \bibinfo {author} {\bibfnamefont {B.~R.}\ \bibnamefont {Ortiz}},
  \bibinfo {author} {\bibfnamefont {S.~D.}\ \bibnamefont {Wilson}}, \bibinfo
  {author} {\bibfnamefont {M.}~\bibnamefont {Dressel}},\ and\ \bibinfo {author}
  {\bibfnamefont {A.~A.}\ \bibnamefont {Tsirlin}},\ }\bibfield  {title}
  {\bibinfo {title} {{Optical detection of the density-wave instability in the
  kagome metal KV3Sb5}},\ }\href {https://doi.org/10.1038/s41535-021-00420-8}
  {\bibfield  {journal} {\bibinfo  {journal} {npj Quantum Materials}\ }\textbf
  {\bibinfo {volume} {7}},\ \bibinfo {pages} {16} (\bibinfo {year}
  {2022})}\BibitemShut {NoStop}%
\bibitem [{\citenamefont {Delacrétaz}\ \emph {et~al.}(2017)\citenamefont
  {Delacrétaz}, \citenamefont {Goutéraux}, \citenamefont {Hartnoll},\ and\
  \citenamefont {Karlsson}}]{luca2017}%
  \BibitemOpen
  \bibfield  {author} {\bibinfo {author} {\bibfnamefont {L.~V.}\ \bibnamefont
  {Delacrétaz}}, \bibinfo {author} {\bibfnamefont {B.}~\bibnamefont
  {Goutéraux}}, \bibinfo {author} {\bibfnamefont {S.~A.}\ \bibnamefont
  {Hartnoll}},\ and\ \bibinfo {author} {\bibfnamefont {A.}~\bibnamefont
  {Karlsson}},\ }\bibfield  {title} {\bibinfo {title} {{Bad Metals from
  Fluctuating Density Waves}},\ }\href
  {https://doi.org/10.21468/SciPostPhys.3.3.025} {\bibfield  {journal}
  {\bibinfo  {journal} {SciPost Phys.}\ }\textbf {\bibinfo {volume} {3}},\
  \bibinfo {pages} {025} (\bibinfo {year} {2017})}\BibitemShut {NoStop}%
\bibitem [{\citenamefont {Fratini}\ and\ \citenamefont
  {Ciuchi}(2021)}]{fratini2021}%
  \BibitemOpen
  \bibfield  {author} {\bibinfo {author} {\bibfnamefont {S.}~\bibnamefont
  {Fratini}}\ and\ \bibinfo {author} {\bibfnamefont {S.}~\bibnamefont
  {Ciuchi}},\ }\bibfield  {title} {\bibinfo {title} {{Displaced Drude peak and
  bad metal from the interaction with slow fluctuations.}},\ }\href
  {https://doi.org/10.21468/SciPostPhys.11.2.039} {\bibfield  {journal}
  {\bibinfo  {journal} {SciPost Phys.}\ }\textbf {\bibinfo {volume} {11}},\
  \bibinfo {pages} {039} (\bibinfo {year} {2021})}\BibitemShut {NoStop}%
\bibitem [{\citenamefont {Rammal}\ \emph {et~al.}(2023)\citenamefont {Rammal},
  \citenamefont {Ralko}, \citenamefont {Ciuchi},\ and\ \citenamefont
  {Fratini}}]{rammal2023}%
  \BibitemOpen
  \bibfield  {author} {\bibinfo {author} {\bibfnamefont {H.}~\bibnamefont
  {Rammal}}, \bibinfo {author} {\bibfnamefont {A.}~\bibnamefont {Ralko}},
  \bibinfo {author} {\bibfnamefont {S.}~\bibnamefont {Ciuchi}},\ and\ \bibinfo
  {author} {\bibfnamefont {S.}~\bibnamefont {Fratini}},\ }\href@noop {}
  {\bibinfo {title} {{Transient localization from the interaction with quantum
  bosons}}} (\bibinfo {year} {2023}),\ \Eprint
  {https://arxiv.org/abs/2312.03840} {arXiv:2312.03840 [cond-mat.str-el]}
  \BibitemShut {NoStop}%
\bibitem [{\citenamefont {Kaufmann}\ and\ \citenamefont
  {Held}(2023)}]{ana_cont}%
  \BibitemOpen
  \bibfield  {author} {\bibinfo {author} {\bibfnamefont {J.}~\bibnamefont
  {Kaufmann}}\ and\ \bibinfo {author} {\bibfnamefont {K.}~\bibnamefont
  {Held}},\ }\bibfield  {title} {\bibinfo {title} {{ana\_cont: Python package
  for analytic continuation}},\ }\href
  {https://doi.org/https://doi.org/10.1016/j.cpc.2022.108519} {\bibfield
  {journal} {\bibinfo  {journal} {Computer Physics Communications}\ }\textbf
  {\bibinfo {volume} {282}},\ \bibinfo {pages} {108519} (\bibinfo {year}
  {2023})}\BibitemShut {NoStop}%
\bibitem [{\citenamefont {Georges}\ \emph {et~al.}(1996)\citenamefont
  {Georges}, \citenamefont {Kotliar}, \citenamefont {Krauth},\ and\
  \citenamefont {Rozenberg}}]{georges1996}%
  \BibitemOpen
  \bibfield  {author} {\bibinfo {author} {\bibfnamefont {A.}~\bibnamefont
  {Georges}}, \bibinfo {author} {\bibfnamefont {G.}~\bibnamefont {Kotliar}},
  \bibinfo {author} {\bibfnamefont {W.}~\bibnamefont {Krauth}},\ and\ \bibinfo
  {author} {\bibfnamefont {M.~J.}\ \bibnamefont {Rozenberg}},\ }\bibfield
  {title} {\bibinfo {title} {{Dynamical mean-field theory of strongly
  correlated fermion systems and the limit of infinite dimensions}},\ }\href
  {https://doi.org/10.1103/RevModPhys.68.13} {\bibfield  {journal} {\bibinfo
  {journal} {Rev. Mod. Phys.}\ }\textbf {\bibinfo {volume} {68}},\ \bibinfo
  {pages} {13} (\bibinfo {year} {1996})}\BibitemShut {NoStop}%
\bibitem [{\citenamefont {Kajueter}\ and\ \citenamefont
  {Kotliar}(1996)}]{kajueter1996}%
  \BibitemOpen
  \bibfield  {author} {\bibinfo {author} {\bibfnamefont {H.}~\bibnamefont
  {Kajueter}}\ and\ \bibinfo {author} {\bibfnamefont {G.}~\bibnamefont
  {Kotliar}},\ }\bibfield  {title} {\bibinfo {title} {{New Iterative
  Perturbation Scheme for Lattice Models with Arbitrary Filling}},\ }\href
  {https://doi.org/10.1103/PhysRevLett.77.131} {\bibfield  {journal} {\bibinfo
  {journal} {Phys. Rev. Lett.}\ }\textbf {\bibinfo {volume} {77}},\ \bibinfo
  {pages} {131} (\bibinfo {year} {1996})}\BibitemShut {NoStop}%
\bibitem [{\citenamefont {Coleman}(2015)}]{coleman2015}%
  \BibitemOpen
  \bibfield  {author} {\bibinfo {author} {\bibfnamefont {P.}~\bibnamefont
  {Coleman}},\ }\href {https://doi.org/10.1017/CBO9781139020916} {\emph
  {\bibinfo {title} {{Introduction to Many-Body Physics}}}}\ (\bibinfo
  {publisher} {Cambridge University Press},\ \bibinfo {year}
  {2015})\BibitemShut {NoStop}%
\bibitem [{\citenamefont {Rohringer}\ \emph {et~al.}(2018)\citenamefont
  {Rohringer}, \citenamefont {Hafermann}, \citenamefont {Toschi}, \citenamefont
  {Katanin}, \citenamefont {Antipov}, \citenamefont {Katsnelson}, \citenamefont
  {Lichtenstein}, \citenamefont {Rubtsov},\ and\ \citenamefont
  {Held}}]{rohringer2018}%
  \BibitemOpen
  \bibfield  {author} {\bibinfo {author} {\bibfnamefont {G.}~\bibnamefont
  {Rohringer}}, \bibinfo {author} {\bibfnamefont {H.}~\bibnamefont
  {Hafermann}}, \bibinfo {author} {\bibfnamefont {A.}~\bibnamefont {Toschi}},
  \bibinfo {author} {\bibfnamefont {A.~A.}\ \bibnamefont {Katanin}}, \bibinfo
  {author} {\bibfnamefont {A.~E.}\ \bibnamefont {Antipov}}, \bibinfo {author}
  {\bibfnamefont {M.~I.}\ \bibnamefont {Katsnelson}}, \bibinfo {author}
  {\bibfnamefont {A.~I.}\ \bibnamefont {Lichtenstein}}, \bibinfo {author}
  {\bibfnamefont {A.~N.}\ \bibnamefont {Rubtsov}},\ and\ \bibinfo {author}
  {\bibfnamefont {K.}~\bibnamefont {Held}},\ }\bibfield  {title} {\bibinfo
  {title} {{Diagrammatic routes to nonlocal correlations beyond dynamical mean
  field theory}},\ }\href {https://doi.org/10.1103/RevModPhys.90.025003}
  {\bibfield  {journal} {\bibinfo  {journal} {Rev. Mod. Phys.}\ }\textbf
  {\bibinfo {volume} {90}},\ \bibinfo {pages} {025003} (\bibinfo {year}
  {2018})}\BibitemShut {NoStop}%
\bibitem [{\citenamefont {Mermin}\ and\ \citenamefont
  {Wagner}(1966)}]{mermin1966}%
  \BibitemOpen
  \bibfield  {author} {\bibinfo {author} {\bibfnamefont {N.~D.}\ \bibnamefont
  {Mermin}}\ and\ \bibinfo {author} {\bibfnamefont {H.}~\bibnamefont
  {Wagner}},\ }\bibfield  {title} {\bibinfo {title} {{Absence of Ferromagnetism
  or Antiferromagnetism in One- or Two-Dimensional Isotropic Heisenberg
  Models}},\ }\href {https://doi.org/10.1103/PhysRevLett.17.1133} {\bibfield
  {journal} {\bibinfo  {journal} {Phys. Rev. Lett.}\ }\textbf {\bibinfo
  {volume} {17}},\ \bibinfo {pages} {1133} (\bibinfo {year}
  {1966})}\BibitemShut {NoStop}%
\bibitem [{\citenamefont {Hertz}(1976)}]{hertz1976}%
  \BibitemOpen
  \bibfield  {author} {\bibinfo {author} {\bibfnamefont {J.~A.}\ \bibnamefont
  {Hertz}},\ }\bibfield  {title} {\bibinfo {title} {{Quantum critical
  phenomena}},\ }\href {https://doi.org/10.1103/PhysRevB.14.1165} {\bibfield
  {journal} {\bibinfo  {journal} {Phys. Rev. B}\ }\textbf {\bibinfo {volume}
  {14}},\ \bibinfo {pages} {1165} (\bibinfo {year} {1976})}\BibitemShut
  {NoStop}%
\bibitem [{\citenamefont {Millis}\ \emph {et~al.}(1990)\citenamefont {Millis},
  \citenamefont {Monien},\ and\ \citenamefont {Pines}}]{millis1990}%
  \BibitemOpen
  \bibfield  {author} {\bibinfo {author} {\bibfnamefont {A.~J.}\ \bibnamefont
  {Millis}}, \bibinfo {author} {\bibfnamefont {H.}~\bibnamefont {Monien}},\
  and\ \bibinfo {author} {\bibfnamefont {D.}~\bibnamefont {Pines}},\ }\bibfield
   {title} {\bibinfo {title} {{Phenomenological model of nuclear relaxation in
  the normal state of
  ${\mathrm{YBa}}_{2}$${\mathrm{Cu}}_{3}$${\mathrm{O}}_{7}$}},\ }\href
  {https://doi.org/10.1103/PhysRevB.42.167} {\bibfield  {journal} {\bibinfo
  {journal} {Phys. Rev. B}\ }\textbf {\bibinfo {volume} {42}},\ \bibinfo
  {pages} {167} (\bibinfo {year} {1990})}\BibitemShut {NoStop}%
\bibitem [{\citenamefont {L\"ohneysen}\ \emph {et~al.}(2007)\citenamefont
  {L\"ohneysen}, \citenamefont {Rosch}, \citenamefont {Vojta},\ and\
  \citenamefont {W\"olfle}}]{lohnezsen2007}%
  \BibitemOpen
  \bibfield  {author} {\bibinfo {author} {\bibfnamefont {H.~v.}\ \bibnamefont
  {L\"ohneysen}}, \bibinfo {author} {\bibfnamefont {A.}~\bibnamefont {Rosch}},
  \bibinfo {author} {\bibfnamefont {M.}~\bibnamefont {Vojta}},\ and\ \bibinfo
  {author} {\bibfnamefont {P.}~\bibnamefont {W\"olfle}},\ }\bibfield  {title}
  {\bibinfo {title} {{Fermi-liquid instabilities at magnetic quantum phase
  transitions}},\ }\href {https://doi.org/10.1103/RevModPhys.79.1015}
  {\bibfield  {journal} {\bibinfo  {journal} {Rev. Mod. Phys.}\ }\textbf
  {\bibinfo {volume} {79}},\ \bibinfo {pages} {1015} (\bibinfo {year}
  {2007})}\BibitemShut {NoStop}%
\bibitem [{\citenamefont {Sch\"afer}\ \emph {et~al.}(2021)\citenamefont
  {Sch\"afer}, \citenamefont {Wentzell}, \citenamefont {\ifmmode~\check{S}\else
  \v{S}\fi{}imkovic}, \citenamefont {He}, \citenamefont {Hille}, \citenamefont
  {Klett}, \citenamefont {Eckhardt}, \citenamefont {Arzhang}, \citenamefont
  {Harkov}, \citenamefont {Le~R\'egent}, \citenamefont {Kirsch}, \citenamefont
  {Wang}, \citenamefont {Kim}, \citenamefont {Kozik}, \citenamefont {Stepanov},
  \citenamefont {Kauch}, \citenamefont {Andergassen}, \citenamefont {Hansmann},
  \citenamefont {Rohe}, \citenamefont {Vilk}, \citenamefont {LeBlanc},
  \citenamefont {Zhang}, \citenamefont {Tremblay}, \citenamefont {Ferrero},
  \citenamefont {Parcollet},\ and\ \citenamefont {Georges}}]{schafer2021}%
  \BibitemOpen
  \bibfield  {author} {\bibinfo {author} {\bibfnamefont {T.}~\bibnamefont
  {Sch\"afer}}, \bibinfo {author} {\bibfnamefont {N.}~\bibnamefont {Wentzell}},
  \bibinfo {author} {\bibfnamefont {F.}~\bibnamefont {\ifmmode~\check{S}\else
  \v{S}\fi{}imkovic}}, \bibinfo {author} {\bibfnamefont {Y.-Y.}\ \bibnamefont
  {He}}, \bibinfo {author} {\bibfnamefont {C.}~\bibnamefont {Hille}}, \bibinfo
  {author} {\bibfnamefont {M.}~\bibnamefont {Klett}}, \bibinfo {author}
  {\bibfnamefont {C.~J.}\ \bibnamefont {Eckhardt}}, \bibinfo {author}
  {\bibfnamefont {B.}~\bibnamefont {Arzhang}}, \bibinfo {author} {\bibfnamefont
  {V.}~\bibnamefont {Harkov}}, \bibinfo {author} {\bibfnamefont {F.~m. c.-M.}\
  \bibnamefont {Le~R\'egent}}, \bibinfo {author} {\bibfnamefont
  {A.}~\bibnamefont {Kirsch}}, \bibinfo {author} {\bibfnamefont
  {Y.}~\bibnamefont {Wang}}, \bibinfo {author} {\bibfnamefont {A.~J.}\
  \bibnamefont {Kim}}, \bibinfo {author} {\bibfnamefont {E.}~\bibnamefont
  {Kozik}}, \bibinfo {author} {\bibfnamefont {E.~A.}\ \bibnamefont {Stepanov}},
  \bibinfo {author} {\bibfnamefont {A.}~\bibnamefont {Kauch}}, \bibinfo
  {author} {\bibfnamefont {S.}~\bibnamefont {Andergassen}}, \bibinfo {author}
  {\bibfnamefont {P.}~\bibnamefont {Hansmann}}, \bibinfo {author}
  {\bibfnamefont {D.}~\bibnamefont {Rohe}}, \bibinfo {author} {\bibfnamefont
  {Y.~M.}\ \bibnamefont {Vilk}}, \bibinfo {author} {\bibfnamefont {J.~P.~F.}\
  \bibnamefont {LeBlanc}}, \bibinfo {author} {\bibfnamefont {S.}~\bibnamefont
  {Zhang}}, \bibinfo {author} {\bibfnamefont {A.-M.~S.}\ \bibnamefont
  {Tremblay}}, \bibinfo {author} {\bibfnamefont {M.}~\bibnamefont {Ferrero}},
  \bibinfo {author} {\bibfnamefont {O.}~\bibnamefont {Parcollet}},\ and\
  \bibinfo {author} {\bibfnamefont {A.}~\bibnamefont {Georges}},\ }\bibfield
  {title} {\bibinfo {title} {{Tracking the Footprints of Spin Fluctuations: A
  MultiMethod, MultiMessenger Study of the Two-Dimensional Hubbard Model}},\
  }\href {https://doi.org/10.1103/PhysRevX.11.011058} {\bibfield  {journal}
  {\bibinfo  {journal} {Phys. Rev. X}\ }\textbf {\bibinfo {volume} {11}},\
  \bibinfo {pages} {011058} (\bibinfo {year} {2021})}\BibitemShut {NoStop}%
\bibitem [{\citenamefont {Johnson}(2005)}]{cubature}%
  \BibitemOpen
  \bibfield  {author} {\bibinfo {author} {\bibfnamefont {S.~G.}\ \bibnamefont
  {Johnson}},\ }\href@noop {} {\bibinfo {title} {{Multi-dimensional adaptive
  integration in {C}: The {Cubature} package}}},\ \bibinfo {howpublished}
  {\url{https://github.com/stevengj/cubature}} (\bibinfo {year}
  {2005})\BibitemShut {NoStop}%
\end{thebibliography}%

\end{document}